 \documentclass[final,1p,times]{elsarticle}

\usepackage{graphicx}
\usepackage{amssymb}

\usepackage{lineno}

\usepackage{xcolor}

\journal{Quantitative Science Studies}

\begin{document}

\begin{frontmatter}



\title{Impact Factor volatility to a single paper: A comprehensive analysis}


\ead{ma2529@columbia.edu}

 \author[label1,label2]{Manolis Antonoyiannakis}
 \address[label1]{Department of Applied Physics \& Applied Mathematics, Columbia University, 500 W. 120th St., Mudd 200, New York, NY 10027}
 \address[label2]{American Physical Society, Editorial Office, 1 Research Road, Ridge, NY 11961-2701
 
 ORCID: 0000-0001-6174-0668}

\begin{abstract}

We study how a single paper affects the Impact Factor (IF) by analyzing data from 3,088,511 papers published in 11639 journals in the 2017 Journal Citation Reports of Clarivate Analytics. We find that IFs are highly volatile. For example, the top-cited paper of 381 journals caused their IF to increase by more than 0.5 points, while for 818 journals the relative increase exceeded 25\%. And one in 10 journals had their IF boosted by more than 50\% by their top {\it three} cited papers. Because the single-paper effect on the IF is inversely proportional to journal size, small journals are rewarded much more strongly than large journals for a highly-cited paper, while they are penalized more for a low-cited paper, especially if their IF is high. This skewed reward mechanism incentivizes high-IF journals to stay small, to remain competitive in rankings. We discuss the implications for breakthrough papers appearing in prestigious journals. We question the reliability of IF rankings given the high IF sensitivity to a few papers for thousands of journals. 

\end{abstract}


\end{frontmatter}


\section{Introduction \& Motivation}
\label{intro}

The effect of a journal's scale (i.e., size) on its citation average cannot be overstated. Recently, we showed (Antonoyiannakis, 2018) that citation averages, such as IFs, are scale-dependent in a way that drastically affects their rankings, and which can be understood and quantified via the Central Limit Theorem: 
For a randomly formed journal of scale $n$, the range of its IF values (measured from the global citation average) scales as $1/\sqrt{n}$. While actual journals are not completely random, the Central Limit Theorem explains to a large extent their IF--scale behavior, and allows us to understand how the balance in IF rankings is tipped in two important ways: (a) Only small journals can score a high IF; and (b) large journals have IFs that asymptotically approach the global citation average as their size increases, via regression to the mean. 

At a less quantitative level, the scale-dependence of citation averages has been noted earlier, by Amin \& Mabe (2004), Campbell (2008), and Antonoyiannakis \& Mitra, (2009). Yet it is almost always neglected in practice. Journal size is thus never {\it accounted} or {\it controlled} for in Impact Factor rankings, whether by {\it standardization} of citation averages (a rigorous approach; see Antonoyiannakis, 2018), or by the grouping of journals in {\it size categories}, much like their grouping in subject categories, a procedure widely accepted due to different citation practices across subjects of research. Instead, Impact Factors for journals of all sizes are lumped together in  rankings such as the Journal Citation Reports (JCR) or the Impact Factor Quartile rankings of Clarivate Analytics, or in ad hoc lists of competitively ranked journals used by faculty hiring and tenure committees, etc. The problem spills over into university rankings and even in national rankings of citation indicators, which generally do not control for the size of a  cohort (department, research field, etc.) in comparing citation averages.

Perhaps the best demonstration of how sensitive citation averages are to scale is to study how they are affected by a single paper. Usually, we take it for granted that averages are scale-independent. However, underlying this certainty is the assumption that a sum over $n$ terms grows (more or less) linearly with scale $n$. In most cases, this assumption is valid. But for research papers in scholarly journals the assumption can break down, because the huge variation in annual citations per paper---from 0 to several thousand---can cause the average to spike abruptly and grow non-linearly  when a highly cited paper is published in a small journal. And while this effect dies out with increasing scale, it routinely plagues IF rankings, because most scholarly journals are  small enough that the effect is present. In short, we need to dispel the notion that size-normalization is equivalent to size-independence for citation averages. 

So, how volatile are Impact Factors, and other citation averages in general? A single research article can tip the balance in university rankings (Waltman et al., 2011; Bornmann \& Marx, 2013) and even affect national citation indicators (Aksnes \& Sivertsen, 2004) when citation averages are used, due to the skewed nature of citation distributions. 
It is also known that in extreme situations, a single paper can strongly boost a journal's IF (Rossner, Van Epps \& Hill, 2007; Dimitrov, Kaveri \& Bayry, 2010; Moed et al., 2012; Foo, 2013; Milojevi{\'c}, Radicchi \& Bar-Ilan, 2017). More recently, Liu et al. (2018) studied the effect of a highly-cited paper on the IF of four different-sized journals in particle physics and found that ``the IFs of low IF and small-sized journals can be boosted greatly from both the absolute and relative perspectives.'' While cautionary remarks have been raised recently at the assumption of size-independence of citation averages (Antonoyiannakis, 2018; Leydesdorff, Bornmann \& Adams, 2019; Lyu \& Shi, 2019; Prathap, 2019; Cope \& Kalantzis, 2014), the overwhelming majority of researchers, bibliometricians, administrators, publishers, and editors continue to use them without realizing or acknowledging the problem. 

In this paper, we show how pervasive  the scale sensitivity of citation averages is, by analyzing the volatility of Impact Factors to a single paper for all 11639 journals listed in the 2017 JCR. 
Our paper is structured as follows. {\it First}, we introduce the volatility index as the IF change, $\Delta f(c)$---or relative change, $\Delta f_r(c)$---when a single paper cited $c$ times is published by a journal of Impact Factor $f$ and size $N$. {\it Second}, we study theoretically how $\Delta f(c)$ depends on $c$, $f$, and $N$, and obtain analytic expressions for the volatility in the general case but also for two special cases: when the new paper is cited well-above or well-below the journal average. We discuss the implications for editorial decisions from the perspective of improving a journal's position in IF rankings. 
{\it Third}, we analyze data from the 11639 journals in the 2017 JCR. We provide summary statistics for the journals' IF volatility to their own top-cited paper. 
We discuss the implications for publishing breakthrough papers in high-profile journals. We also discuss the reliability of IF rankings, and provide recommendations for more meaningful and statistically viable comparisons of journals' citation impact.

The high volatility values from real-journal data demonstrate that ranking journals by IFs constitutes a non-level playing field, since, depending on size, the IF gain of publishing an equally cited paper can span up to 4 orders of magnitude across journals. It is therefore critical to consider novel ways of comparing journals based on solid statistical grounds. 

\section{How a single paper affects the IF: An example from four journals}
\label{2.1}

We are now ready to analyze the effect of a single paper on a journal's IF. 
Initially, let the journal have an IF equal to $f_1$, which is the ratio of $C_1$ citations to the biennial publication count $N_1$. The additional paper causes the IF denominator to increase by 1, and the nominator by $c$. 

Before we study the general case, let us first consider one example, to get an idea of what is going on. In Table 1 we list four journals whose sizes range from 50 to 50,000, but their IFs are the same. The numbers are fictitious but realistic: As one can confirm from the JCR, there are journals with size and IFs sufficiently close to the values in the table. 
\begin{table}[h]
\centering
\begin{tabular}{l l l l l l l l}
\hline
Journal & Size & Citations & Initial IF & New paper & Final IF & $\Delta f(c)$ & $\Delta f_r(c)$  \\
 & $N_1$ & $C_1$ & $f_1$ & $c$ & $f_2$ & $f_2-f_1$  & $(f_2-f_1)/f_1$ \\
\hline
 A & 50 & 150 & 3 & 100 & 4.902 & 1.902 & 63.4 \% \\
 B & 500 & 1500 & 3 & 100 & 3.194 & 0.194 & 6.45 \% \\
 C & 5,000 & 15,000 & 3 & 100 & 3.019 & 0.019 & 0.65 \% \\
 D & 50,000 & 150,000 & 3 & 100 & 3.002 & 0.0019 & 0.06 \% \\
\hline
\end{tabular}
\caption{A hypothetical but realistic scenario. Four journals, {\it A, B, C}, and {\it D}, have the same IF but different sizes, when they each publish a paper that brings them $c=100$ citations. The IF gain spans 4 orders of magnitude---both in absolute, $\Delta f(c)$, and relative, $\Delta f_r(c)$, terms---since it depends not only on the additional paper, but also on the size of each journal.}
\end{table}

Journal {\it B} is 10 times larger than {\it A}. When a highly-cited paper ($c = 100$) is published by {\it A}, the IF changes by $\Delta f(100) = 1.902$. When the same paper is published by {\it B}, the change is ten times smaller, i.e., $\Delta f(100) = 0.194$. Therefore, to compete with journal {\it A}---to obtain the same IF increase $\Delta f(c)$---journal {\it B} needs to publish 10 equally highly cited papers. Likewise, for every  paper of $c=100$ that {\it A} publishes, {\it C} needs to publish 100 equally cited papers to obtain the same IF increase. And for every paper of $c=100$ that journal {\it A} publishes, journal {\it D} needs to publish 1000 equally cited papers to compete.

To sum up, the IF increase  is inversely proportional to journal size. Publication of the same highly cited paper in any of the journals {\it A, B, C,} or {\it D},  produces widely disparate outcomes, as the corresponding IF increase spans four orders of magnitude, from 0.0019 to 1.902. 
With such a high sensitivity to scale, the comparison of IFs of these four journals is no level playing field:  Small journals disproportionately benefit from highly cited papers. 

The above example considers a highly cited paper. As we will shortly see, there is a sufficient number of highly cited papers to cause hundreds of journals every year to jump up considerably in IF rankings due to one paper. And even further: There are many journals of sufficiently small size {\it and} small IF that even a low- or moderately-cited paper can produce a big increase in their IF. Therefore, IF volatility due to a single paper (or a handful of papers, in the more general case) is a much more common pattern than is widely recognized. Which is why this behavior of IFs goes beyond academic interest. To understand this fully, let us now consider the general case.

\section{How a single paper affects the IF: The general case. Introducing the IF volatility index.}
\label{2.2}

The initial IF is
\begin{equation}
f_1=\frac{C_1}{N_1}, \label{eq:f0-def}
\end{equation}
so that when the new paper is published by the journal, the new IF becomes
\begin{equation}
f_2=\frac{C_1+c}{N_1+1}. \label{eq:f-def}
\end{equation}
The change (volatility) in the IF caused by this one paper is then 
\begin{equation}
\Delta f(c) \equiv f_2-f_1 = \frac{C_1+c}{N_1+1}-\frac{C_1}{N_1}, \label{eq:Delta f-def}
\end{equation}
so that
\begin{equation}
\Delta f(c) = \frac{c-f_1}{N_1+1} \approx \frac{c-f_1}{N_1}, \label{eq:Delta f}
\end{equation}
where the approximation is justified for $N_1 \gg 1$, which applies for all but a few journals that publish only a few items per year. So, the IF volatility, $\Delta f(c)$, depends both on the new paper (i.e., on $c$) and on the journal (size $N_1$, and citation average $f_1$) where it is published. 

We can also consider the {\it relative change} in the citation average caused by a single paper, which is probably a more pertinent measure of volatility.  For example, if a journal's IF jumps from 1 to 2, then this is bigger news than if it jumped from 20 to 21. The relative volatility is
\begin{equation}
\Delta f_r(c) \equiv \frac{f_2-f_1}{f_1} = \frac{c-f_1}{f_1 (N_1+1)} \approx \frac{c-f_1}{f_1 N_1} = \frac{c-f_1}{C_1}, \label{eq:Delta_r}
\end{equation}
where, again, the approximation is justified when $N_1 \gg 1$. The above equation can be further simplified for highly cited papers ($c \gg f_1$) as
\begin{equation}
\Delta f_r(c)  \approx \frac{c}{C_1}, \;\; {\rm when}\; c \gg f_1.  \label{eq:Delta_r.high_c}
\end{equation}

Let us now return to $\Delta f(c)$ and make a few remarks.

(a) For $c>f_1$, the additional paper is above-average with respect to the journal, and there is a {\it benefit} to publication: $\Delta f(c) >0$ and the IF increases, i.e., $f_2>f_1$.

(b) For $c<f_1$, the new paper is below-average with respect to the journal, and publishing it invokes a {\it penalty}: $\Delta f(c) <0$ as the IF drops, i.e., $f_2<f_1$. 

(c) For $c=f_1$, the new paper is average, and publishing it makes no difference in the IF.

(d) The presence of $N_1$ in the denominator means that the benefit or penalty of publishing an additional paper decays rapidly with journal size. This has dramatic consequences. 
 
Let us now consider two special cases of interest:

\noindent 
\ding{226}
{\bf CASE 1.} 	The new paper is well above average relative to the journal, i.e., $c \gg f_1$. Here,
\begin{equation}
\Delta f(c) = \frac{c-f_1}{N_1+1} \approx \frac{c}{N_1+1} \approx \frac{c}{N_1}, \label{eq:c>>f0}
\end{equation}
where the last step is justified since in realistic cases we have $N_1 \gg 1$. The volatility $\Delta f(c)$ depends on the paper itself and on the journal size. The presence of $N_1$ in the denominator means that publishing an above-average paper is far more beneficial to small journals than to large journals. For example, a journal {\it A} that is ten times smaller than a journal {\it B} will have a ten times higher benefit upon publishing the same highly cited paper, even if both journals had the same IF to begin with! The {\it editorial implication} here is that it pays for editors of small journals to be particularly watchful for high-performing papers. From the perspective of competing in IF rankings, small journals have two conflicting incentives: Be open to publishing risky and potentially breakthrough papers on the one hand, but not publish too many papers lest they lose their competitive advantage due to their small size. 

For $c\ll N_1$, we get $\Delta f(c) \approx 0$ even for large $c$. So, when large journals publish highly cited papers, they have a tiny benefit in their IF. For example, when a journal with $N_1 = 2000$ publishes a paper of $c = 100$, the benefit is a mere $\Delta f(100) = 0.05$. For a very large journal of $N_1 = 20000$, even an extremely highly cited paper of $c = 1000$ produces a small gain $\Delta  f(1000) = 0.05$. 
 
But for small and intermediate values of $N_1$, the value of $\Delta f(c)$ can increase appreciably. This is the most interesting regime for journals, which tend to be rather small: Recall that ``90\% of all journals publish 250 or fewer citable items annually'' (Antonoyiannakis, 2018).    

\noindent 
\ding{226}
{\bf CASE 2.}  The new paper is well below average relative to the journal, i.e., $c \ll f_1$.  (For journals of, say, $f_1 \le 2$, the condition $c \ll f_1$ implies $c=0$.)  Here,
\begin{equation}
\Delta f(c) = \frac{c-f_1}{N_1+1} \approx -\frac{f_1}{N_1+1} \approx -\frac{f_1}{N_1}, \label{eq:c<<f0}
\end{equation}
since in realistic cases we have $N_1 \gg 1$. The penalty $\Delta f(c)$ depends now only on the journal parameters ($N_1, f_1$), and it is greater for small-sized, high-IF journals. The {\it editorial implication} is that editors of small and high-IF journals need to be more vigilant in pruning low-performing papers than editors of large journals. Two kinds of papers are low-cited, at least in the IF citation window: (a) archival, incremental papers, and (b) some truly ground-breaking papers that may appear too speculative at the time and take more than a couple years to be recognized.  

For $f_1 \ll N_1$, we get  $\Delta f(c) \approx 0$. Very large journals lose little by publishing low-cited papers.

The take-home message from the above analysis is two-fold. First, with respect to increasing their IF, it pays for all journals to take risks. Because the maximum penalty for publishing below-average papers ($\approx f_1/N_1$) is smaller than the maximum benefit for publishing above-average papers ($\approx c/N_1$), it is better for a journal's IF that its editors publish a paper they are on the fence about, if what is at stake is the possibility of a highly influential paper. Some of these papers may reap high citations to be worth the risk: recall that $c$ can lie in the hundreds or even thousands. 

However, the reward for publishing breakthrough papers is {\it much} higher for small journals. For a journal's IF to seriously benefit from ground-breaking papers, the journal must above all remain small, otherwise the benefit is much reduced due to its inverse dependence with size. To the extent that editors of elite journals are influenced by IF considerations, they have an incentive to keep a tight lid on their acceptance decisions and reject many good papers, and even some potentially breakthrough papers they might otherwise have published. We wonder whether the abundance of prestigious high-IF journals with small biennial sizes, $N_{2Y} < 400$, and especially their size stability over time, bears any connection to this realization. In other words, is {\it ``size consciousness''} a reason why high-IF journals stay small? We claim {\it yes}.  As Philip Campbell, former Editor-in-Chief of {\it Nature} put it, ``The larger the number of papers, the lower the impact factor. In other words, worrying about maximizing the impact factor turns what many might consider a benefit---i.e. more good papers to read---into a burden." (Campbell, 2008).

To recap, why are high-IF journals incentivized to stay small to remain competitive in IF rankings? Because once a journal reaches a high IF, it is much easier to sustain it by staying small than by expanding in size. Equation \ref{eq:Delta f} explains why. For every above-average paper (of fixed citation count, for simplicity of argument) published, the IF increases but by a smaller amount as the journal grows, so the returns diminish. At the same time, for every below-average paper published the IF drops. With increasing journal size, it gets harder to keep raising the IF but easier to slip into a lower IF, since low-cited papers are far more abundant than highly cited papers. It is a matter of risk.

 The incentive for high-IF journals to stay small may disproportionately affect groundbreaking papers, because they entail higher risk. How so? First, it is hard to identify such papers before publication.  Many groundbreaking papers face controversy in the review process and are misjudged by referees who may be too conservative or entrenched to realize their transformative potential. Obviously, no editor wishes to publish unrealistic or wrong papers. The editors hedge their bets, so to speak, and take chances in accepting controversial papers.  (Needless to say, this is where editorial skill and competence, coupled with outstanding and open-minded refereeing, can make a difference.) But editors of small, high-IF journals can afford fewer risks than editors of large, moderate-IF journals, as explained in the previous paragraph---which pushes them to be more conservative and accept a smaller fraction of these controversial papers. Second, even if it were possible to know the groundbreaking papers  beforehand, editors of small, high-IF journals would still be incentivized by IF arguments to reject some of them, because such papers are less likely to be top-cited in the 2-year IF window. Indeed, Wang, Veugelers \& Stephan (2017) reported on the increased difficulty of transformative papers to appear in prestigious journals. They found that ``novel papers are less likely to be top cited when using short time-windows,'' and ``are published in journals with Impact Factors lower than their non-novel counterparts, {\it ceteris paribus}.'' They argue that the increased pressure on journals to boost their IF ``suggests that journals may strategically choose to not publish novel papers which are less likely to be highly cited in the short run.'' 
To sum up: If a small journal fine-tunes its risk level and publishes only some controversial (i.e., potentially groundbreaking) papers, its IF will benefit more, statistically speaking, than if it published them all.

Why worry about intellectually risky papers? Because they are more likely to lead to major breakthroughs (Fortunato {\it et. al.}, 2018.) It was in this spirit that the {\it Physical Review Letters} Evaluation Committee recommended back in 2004 that steps be taken to ``educate referees to identify cutting edge papers worth publishing even if their correctness cannot be definitively established,'' and that ``[r]eferee training should emphasize that a stronger attempt be made to accept more of the speculative exciting papers that really move science forward.'' (Cornell {\it et al.}, 2004.) Granting agencies have reached a similar understanding. For example, an effort to encourage risk in research is the {\it NIH Common Fund Program}, established in 2004 and supporting ``compelling, high-risk research proposals that may struggle in the traditional peer review process despite their transformative potential'' (NIH News Release, 2018). 
These awards ``recognize and reward investigators who have demonstrated innovation in prior work and provide a mechanism for them to go in entirely new, high-impact research directions.'' (Collins, Wilder \& Zerhouni, 2014).
Europe's flagship program for funding high-risk research, the European Research Council, was established in 2007 and ``target[s] frontier research by encouraging high-risk, high-reward proposals that may revolutionize science and potentially lead to innovation if successful.'' (Antonoyiannakis, Hemmelskamp \& Kafatos, 2009).

\section{How the IF volatility index $\Delta f(c)$ depends on the parameters $f_1, N_1, c$.}
\label{2.3}

We now analyze graphically how $\Delta f(c)$ depends on its parameters, namely, the IF of the journal $f_1$, the biennial publication count $N_1$, and the annual citation count $c$ of a single paper. 

First, let us briefly comment on the dependence of $\Delta f(c)$ on $f_1$. Impact Factors $f_1$ range typically from 0.001--200, but are heavily concentrated in low-to-moderate values (Antonoyiannakis, 2018): The most commonly occurring value (the mode) is 0.5, while 75\% of all journals in the 2017 JCR have ${\rm IF}<2.5$. Since our chief aim here is to study the effects of a single paper on citation averages, we are mostly interested in high $c$ values ($c>100$, say), in which case the effect of $f_1$ on $\Delta f(c)$ or on $\Delta f_r(c)$ can be usually ignored, as can be seen from Eq. (\ref{eq:c>>f0}) and Eq. (\ref{eq:Delta_r.high_c}) respectively. For smaller $c$ values relative to $f_1$, the effect of $f_1$ is to simply reduce the size of $\Delta f(c)$ by some amount, but is otherwise of no particular interest. 

Let us now look at the dependence of $\Delta f(c)$ on $N_1$ and $c$. The journal biennial size $N_1$ ranges from 20--60,000 and is heavily centered at small sizes (Antonoyiannakis, 2018), which has important implications, as we shall see. As for $c$, it ranges from 0--5000 in any JCR year, and its distribution follows a power law characteristic of the Pareto distribution for $c \ge 10$ (Table 2).

\begin{table}[h]
\centering
\begin{tabular}{l l l l l l l l}
\hline
\textbf{Citation} & No. papers cited & \textbf{Citation} & No. papers cited \\
\textbf{threshold, $c_t$} &  {\it at least} $c_t$ times & \textbf{threshold, $c_t$} &  {\it at least} $c_t$ times\\
\hline
0	&	3088511	 & 200	&	302	 \\
1	&	2138249	 & 300	&	139	 \\
2	&	1490683	 & 400	&	80	\\
5	&	570744	& 500	&	56	 \\
10	&	176718	& 1000	&	14	 \\
\hline
20	&	43030  & 1500	&	7 \\
30	&	18485  & 2000	&	6 \\
40	&	10016  & 2500	&	5 \\
50	&	6222   & 3000	&	2 \\
100	&	1383   & 4000	&	0 \\
\hline
\end{tabular}
\caption{Number of papers cited {\it at least} $c_t$ times. Publication Years = 2015--2016, Citation Year = 2017. JCR data.}
\end{table}

Figure \ref{fig:Delta_c.N_Y.c-1} is a 3D surface plot of $\Delta f(c)$ vs. $N_1$ and $c$, for a fixed $f_1=10$. Figure \ref{fig:Delta_c.N_Y.c-2} is projection of Fig. \ref{fig:Delta_c.N_Y.c-1} in 2D, i.e., a contour plot, for more visual clarity. The main features of the plots are:

\begin{figure}
\centering
\includegraphics[width=1\linewidth]{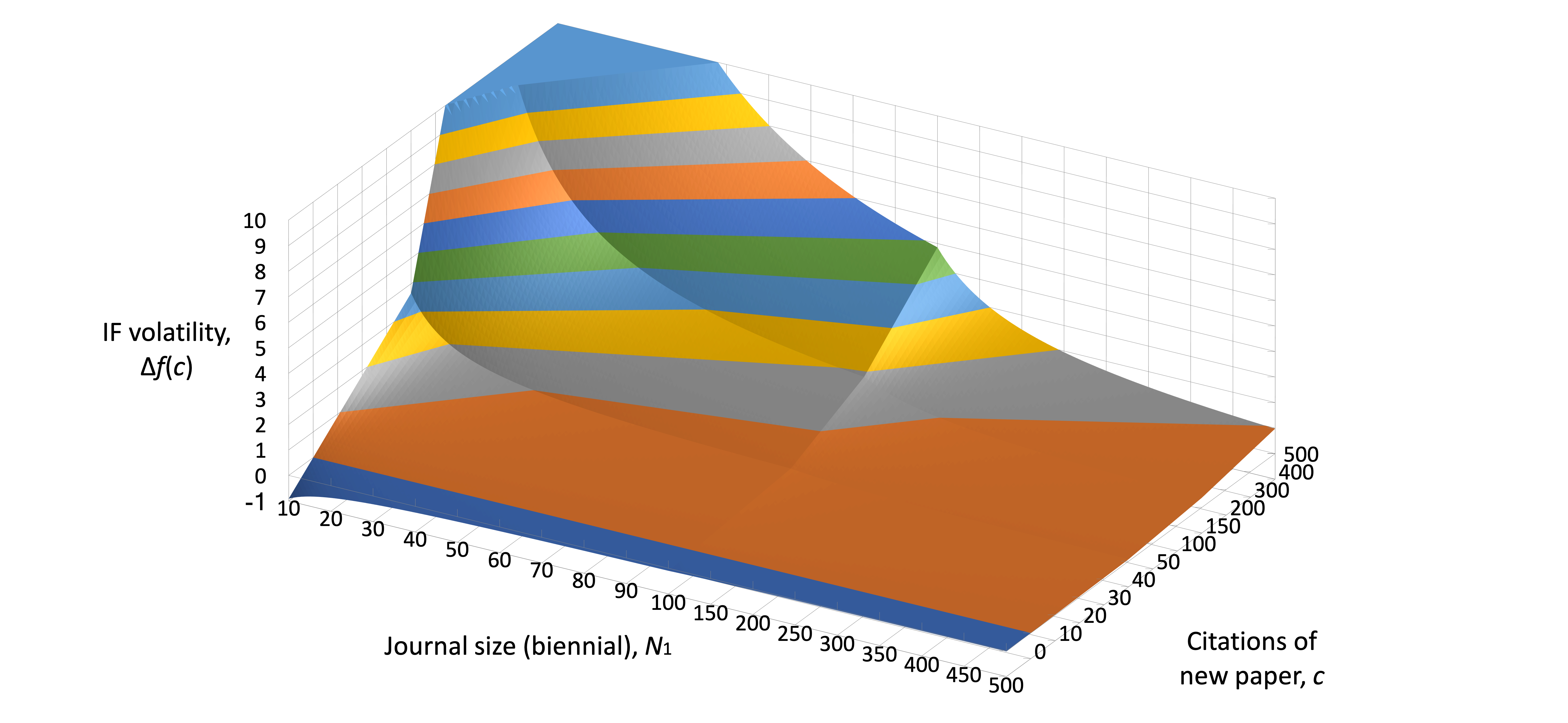}
\caption{3D surface plot of IF volatility $\Delta f(c)$ vs.  (biennial) journal size $N_1$ and citation count $c$ of the new paper, for a journal whose IF was $f_1=10$ before publishing the  paper. The range of $N_1$ values plotted here covers 90\% of all journals, while 50\% of all journals publish $\sim 130$ or fewer citable items biennially (Antonoyiannakis, 2018). So, for thousands of journals a paper cited $c\simeq 100$ can cause $\Delta f(c)>1$. The IF of the journal has little effect on $\Delta f(c)$ as long as $c \gg f_1$. See Eq. (\ref{eq:Delta f}).}
\label{fig:Delta_c.N_Y.c-1}
\end{figure}

\begin{figure}
\centering
\includegraphics[width=.8\linewidth]{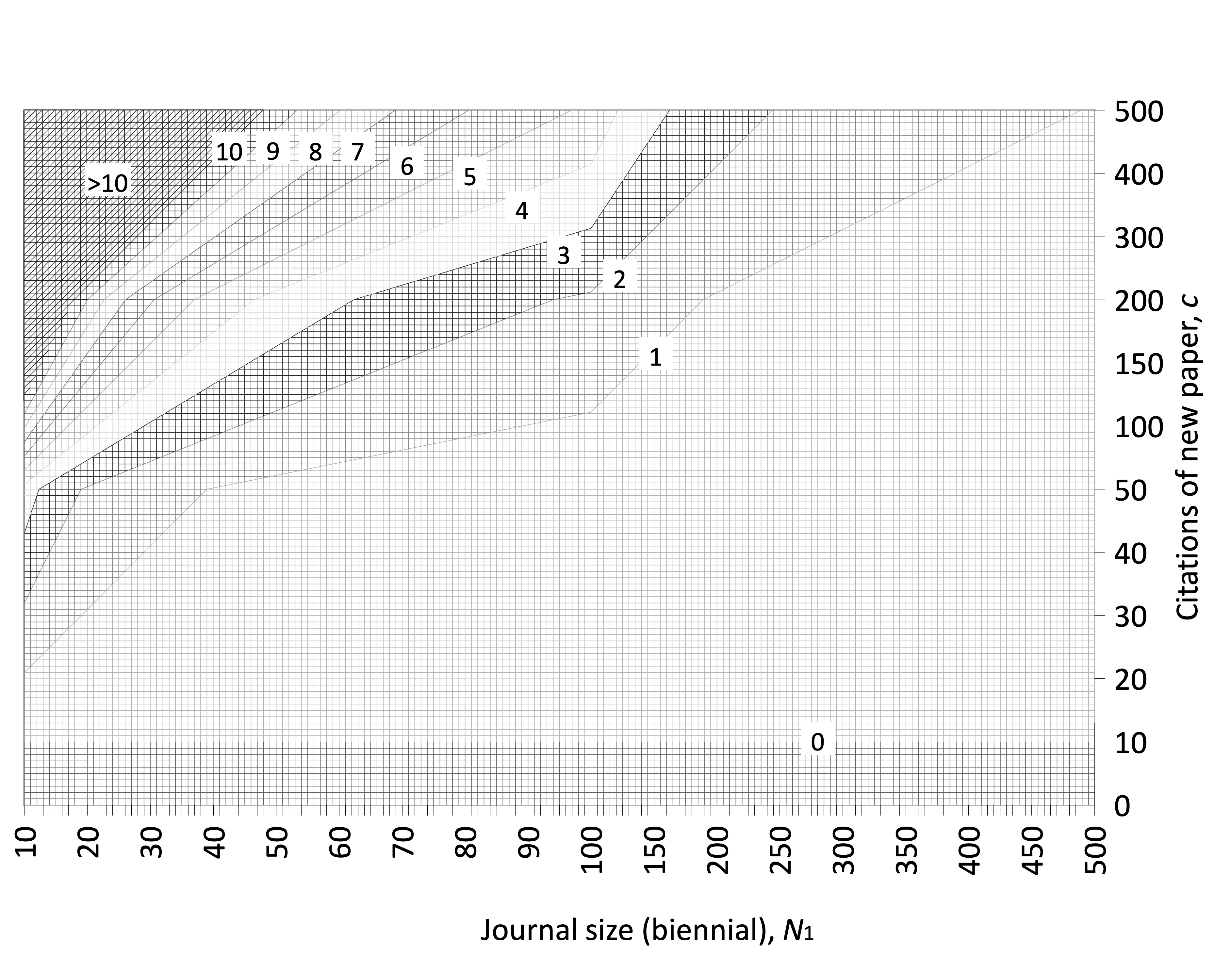}
\caption{Same data as in Fig. \ref{fig:Delta_c.N_Y.c-1} but in a contour plot. Changes in gray level denote crossing integer values of $\Delta f(c)$ as shown in the plot. As journal size decreases, the IF volatility $\Delta f(c)$ increases.}
\label{fig:Delta_c.N_Y.c-2}
\end{figure}

1. For a given $c$ value, $\Delta f(c)$ decreases rapidly with $N_1$, as expected from Eq. (\ref{eq:Delta f}), since the two quantities are essentially inversely proportional for $c \gg f_1$. 

2. For {\it realistic} values of $N_1, c$, the volatility $\Delta f(c)$ can take high values. For example, for $20 \leq N_1 \leq 100$ and $20 \leq c \leq 500$ we have  $0.5 < \Delta f(c) < 25$. Think about it: A single paper can raise the IF of these journals by several points! This is impressive.  

Why are these parameter values realistic? Because small journals abound, while there are thousands of sufficiently cited papers that can cause an IF spike. Indeed, 25\% of the 11639 journals in our data set publish fewer than 68 items biennially ($N_1< 68$), while 50\% of journals publish fewer than 130 items, and 75\% of journals publish fewer than 270 items. The range of $N_1$ values plotted here (10--500) spans 90\% of all journals (Antonoyiannakis, 2018). At the same time, 6222 papers in our data set were cited at least 50 times, 1383 papers were cited at least 100 times, 302 papers were cited at least 200 times, etc. (Table 2). 

As these plots demonstrate, small journals ($N_1 \leq 500$) enjoy a disproportionate benefit upon publishing a highly-cited paper, compared to larger journals. Small journals are abundant. Highly cited papers are relatively scarce, but nevertheless exist in sufficient numbers to cause abrupt IF spikes for hundreds of small journals. 

But an additional effect is also at work here, and it can cause IF spikes for thousands of journals: a medium-cited paper published in a small and otherwise little-cited journal. Given the high abundance of medium-cited papers (e.g., more than 176,000 papers in our data set are cited at least 10 times) {\it and} low-IF journals (e.g., 4046 journals have $f_1 \le 1$), journals that would otherwise have had a negligible IF can end up with small or moderate IF. This is a much more commonly occurring effect than has been realized to date.

\section{Systematic study of the volatility index, using  data from 11,639 journals.}
\label{2.4}

Now that we understand in theory the IF volatility, let us look at some real journal data. We have analyzed all journals listed in the 2017 JCR of Clarivate Analytics. 

At this point, we could continue to study the effect of a hypothetical paper on the IFs of actual journals, using JCR data for IFs and journal sizes. For example, we could ask the question, ``How does the IF of each journal change by incorporation of a paper cited $c=100$ times?'' and calculate the corresponding volatility $\Delta f(100)$. 
While such a calculation would be of value, we adopt a different approach, in order to stay firmly anchored on actual data from both journals {\it and} papers, and avoid hypotheticals. We ask the question ``How did the IF (citation average) of each journal change by incorporation of its most cited paper, which was cited $c^*$ times in the IF 2-year time-window?'' We thus calculate the quantity $\Delta f(c^*)$, where $c^*$ is no longer constant and set equal to some hypothetical value, but varies across journals. 

First, a slight change in terminology to avoid confusion. We wish to study the effect of a journal's top-cited paper on its citation average $f$ when its biennial publication count is $N_{2Y}$. So, our journal's initial state has size $N_1=N_{2Y}-1$ and citation average $f_1$, which we will denote as $f^*$. Our journal's final state has  $N_2=N_{2Y}$ and $f_2=f$, upon publication of the top-cited paper that was cited $c^*$ times. We study how $\Delta f(c^*)$ and $\Delta f_r(c^*)$ behave using JCR data.

\begin{figure}
\centering
\includegraphics[width=0.9\linewidth]
{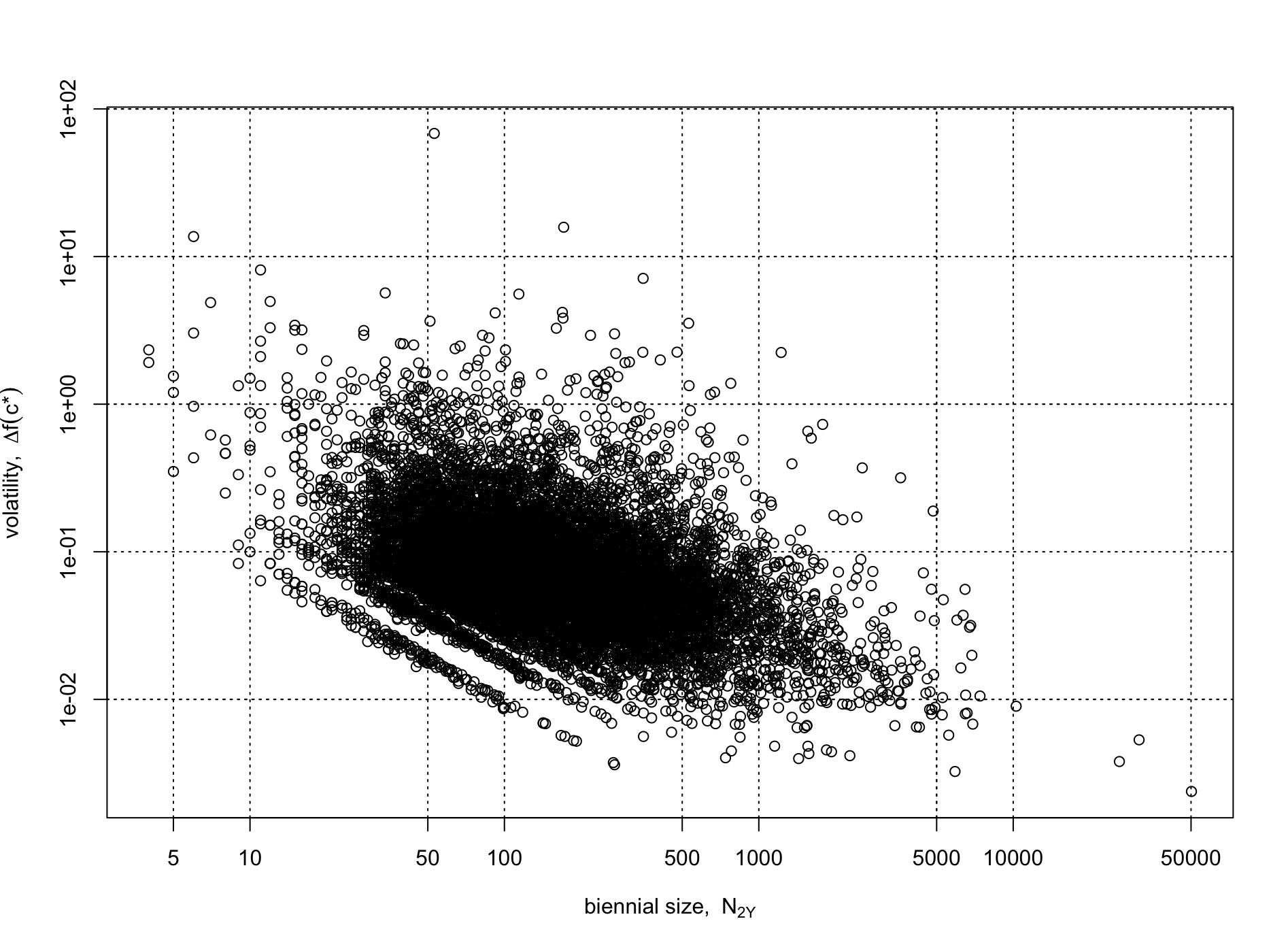}
\caption{IF volatility, $\Delta f(c^*)$, vs. journal (biennial) size, $N_{2Y}$, for 11639 journals in the 2017 JCR. 
}
\label{fig:Delta-N0}
\end{figure}

Now, some technical details. The analysis was carried out in the second half of 2018. Among the 12,266 journals initially listed in the 2017 JCR, we removed the several hundred duplicate entries, as well as the few journals whose IF was listed as zero or not available. We thus ended up with a master list of 11639 unique journal titles that received a 2017 IF as of December 2018. For each journal in the master list we obtained its individual Journal Citation Report, which contained the 2017 citations to each of its citable papers (i.e., articles and reviews) published in 2015--2016. We were thus able to calculate the citation average $f$ for each journal, which approximates the IF and becomes identical to it when there are no ``free'' or ``stray'' citations in the numerator---that is, citations to front-matter items such as editorials, letters to the editor, commentaries, etc., or citations to the journal without specific reference of volume and page or article number. We will thus use the terms ``IF'' and ``citation average'' interchangeably, for simplicity. Collectively, the 11639 journals in our master list published 3,088,511 papers in 2015--2016, which received 9,031,575 citations in 2017 according to the JCR. This is our data set. 

For the record, for 26 journals the top cited paper was the only cited paper, in which case $f^*=0$. Also, for 11 journals none of their papers received any citations, in which case $f=f^*=0$! (These journals were however allocated an IF, so they did receive citations to the journal and year, or to their front matter.) None of these 37 journals is depicted in our log-log plots.

\begin{figure}
\centering
\includegraphics[width=0.9\linewidth]
{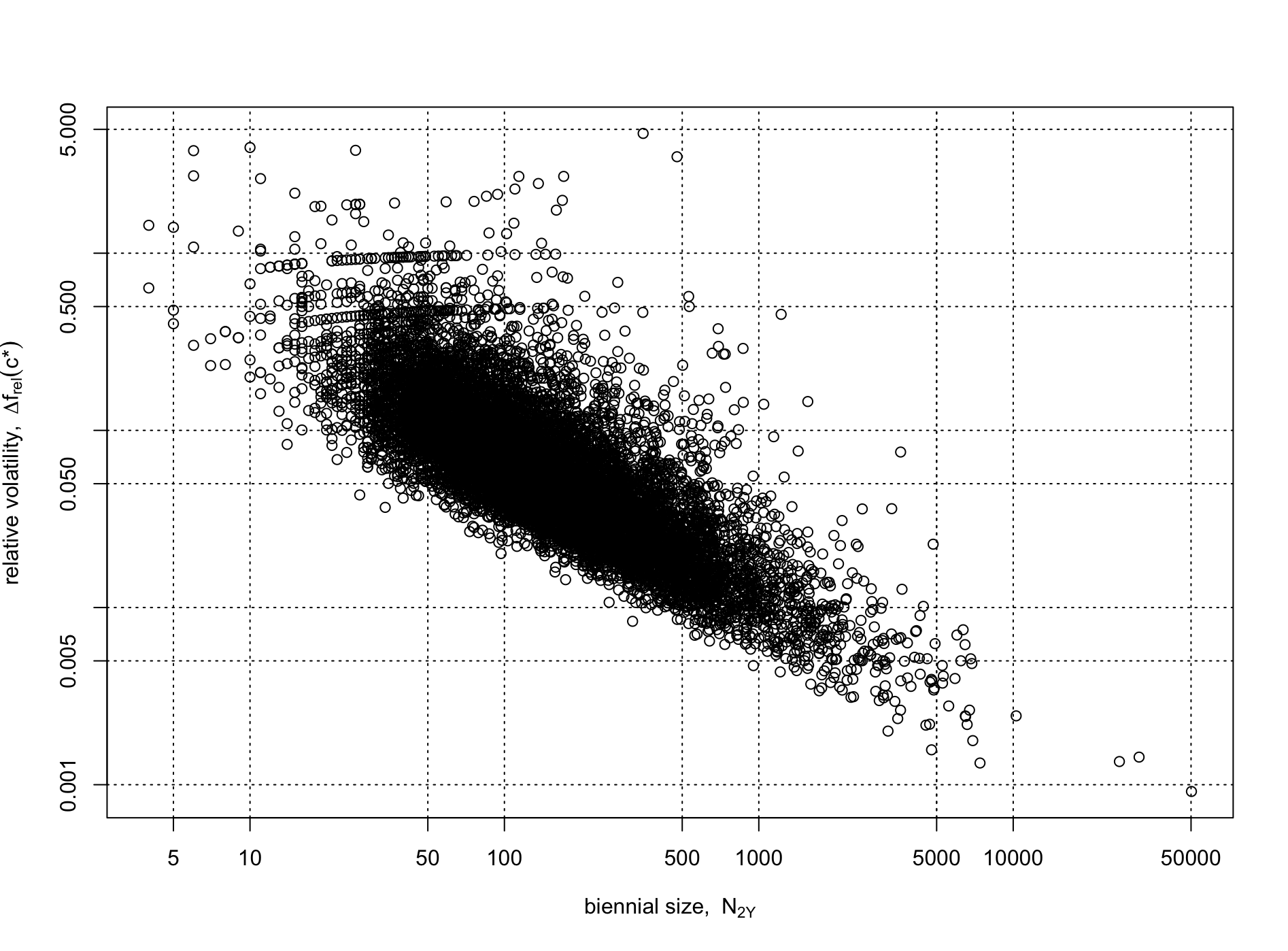}
\caption{IF relative volatility, $\Delta f_r(c^*)$ vs. journal (biennial) size, $N_{2Y}$, for 11639 journals in the 2017 JCR.}
\label{fig:Delta_r-N0}
\end{figure}

In Figs. \ref{fig:Delta-N0} and \ref{fig:Delta_r-N0} we plot the volatility $\Delta f(c^*)$ and relative volatility $\Delta f_r(c^*)$, respectively, vs. journal size $N_{2Y}$. In Fig. \ref{fig:vol-f0} we plot the volatility $\Delta f(c^*)$ vs. the journal citation average, $f$, in a bubble plot where bubble size is proportional to journal size. In Fig. \ref{fig:c*-f0} we plot the citation count of the top-cited paper, $c^*$, vs. journal citation average, $f$. 
In Tables 3 \& 4 and 7 \& 8 we identify the top 100 journals in decreasing volatility $\Delta f(c^*)$ and relative volatility $\Delta f_r(c^*)$, respectively. In Tables 5 and 6 we show the frequency distribution of $\Delta f(c^*)$ and $\Delta f_r(c^*)$, respectively.

Our key findings are as follows.

\begin{enumerate}

\item {\it High volatilities are observed for hundreds of journals.} For example: 

\noindent
(a) $\Delta f(c^*) >0.5$ for 381 journals,

\noindent
(b) $\Delta f(c^*) >0.25$ for 1061 journals, etc.

Relative volatilities are also high for hundreds of journals. For example:

\noindent
(c)  $\Delta f_r(c^*) > 50\%$ for 231 journals,

\noindent
(d) $\Delta f_r(c^*) > 25\%$ for 818 journals, etc.

\item 
If we look at the top {\it few} cited papers per journal---as opposed to the {\it single} top cited paper---then {\it the IF sensitivity to a handful of papers becomes even more dramatic.} For instance, the IF was boosted by more than 50\% by: 

\noindent
(a) the top {\it two} cited papers for 710 journals, 

\noindent
(b) the top {\it three} cited papers for 1292 journals, 

\noindent
(c) the top {\it four} cited papers for 1901 journals, etc.

So, 10\% of journals had their IF boosted by more than 50\% by their top {\it three} cited papers! 

\item {\it Highest volatility values occur for small journals.} This agrees with our earlier finding that smaller journals benefit the most from a highly cited paper. By ``small journals'' we mean $N_{2Y} \le 500$.
For example, 97 of the top 100 journals ranked by volatility (Tables 3 and 4), and {\it all} the top 100 journals ranked by relative volatility (Tables 7 and 8) publish fewer than 500 papers biennially ($N_{2Y} \le 500$).

\item {\it Above the limit of $N_{2Y} \approx 500$, journal size starts to become prohibitively large for a journal's IF to profit from highly cited papers.} Notice how the maximum values of $\Delta f(c^*)$ and $\Delta f_r(c^*)$ follow a downward trend with journal size above this limit.

\begin{figure}
\centering
\includegraphics[width=0.9\linewidth]
{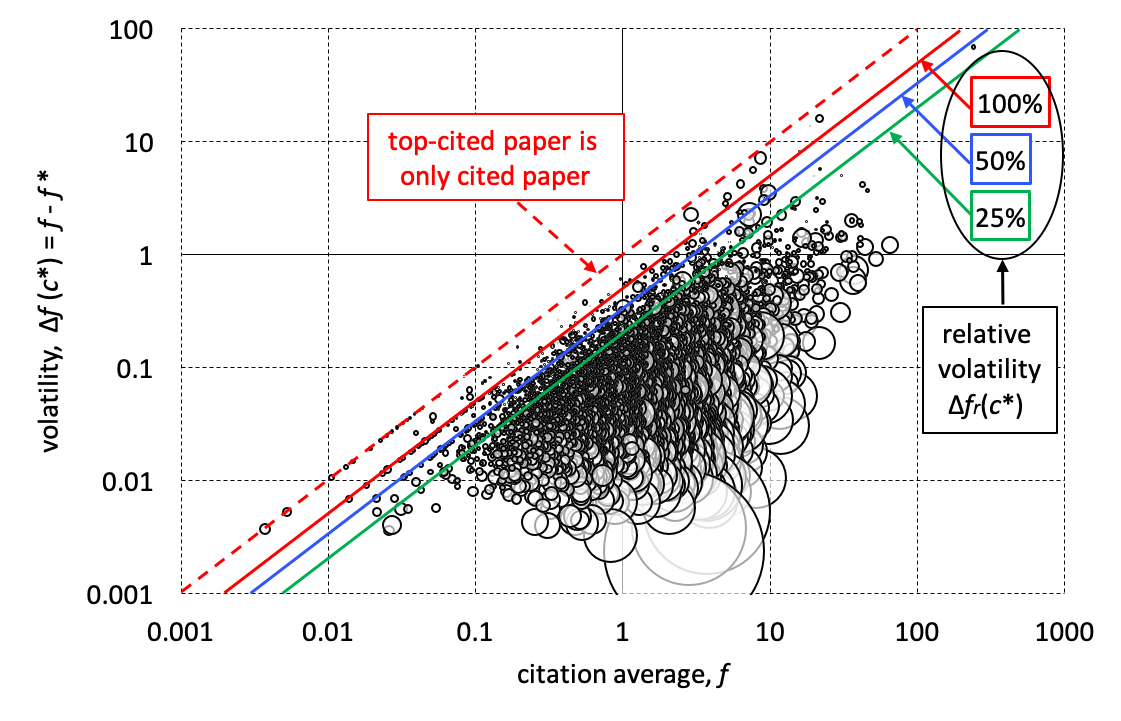}
\caption{IF volatility, $\Delta f(c^*)$, vs. citation average, $f$,  for 11639 journals in the 2017 JCR. Bubble size shows journal size. The dashed line corresponds to the top-cited paper having {\it all} the journal's citations, which occurs for 26 journals. The three parallel lines labeled ``100\%'', ``50\%'', and ``25\%'' denote relative volatility values $\Delta f_r(c^*)$---i.e., relative IF boost---caused by the top-cited paper. Thus, data points above the 25\% line describe the 818 journals whose top-cited paper boosted their IF by more than 25\%. As expected from the Central Limit Theorem, increasing journal size causes the volatility to drop (larger bubbles ``fall'' to the bottom) and the IF to approach the global citation average $\mu=2.9$.} 
\label{fig:vol-f0}
\end{figure}

\item {\it For some journals, an extremely highly cited paper causes a large volatility $\Delta f(c^*)$.} Consider the top 2 journals in Table 3. The journal {\it CA-A Cancer Journal for Clinicians} published in 2016 a paper that was cited 3790 times in 2017, accounting for almost 30\% of its IF  citations that year, with a corresponding $\Delta f(c^*)=68.3$. Without this paper, the journal's citation average would have dropped from $f=240.1$ to $f^*=171.8$. Similarly, the {\it Journal of Statistical Software} published in 2015 a paper cited 2708 times in 2017, capturing 73\% of the journal's citations that year. Without this paper, the journal's citation average would have dropped from $f=21.6$ to $f^*=5.8$. Although such extreme volatility values are rare, they occur every year. 

\item {\it A paper need not be exceptionally cited to produce a large IF boost, provided the journal is sufficiently small.} Consider the journals ranked \#3 and \#4 in Table 3, namely, {\it Living Reviews in Relativity} and {\it Psychological Inquiry}. These journals' IFs were strongly boosted by their top-cited paper, even though the latter was much less cited ($c^*=87$ and 97, respectively) than for the top 2 journals. This is because journal sizes were  smaller also ($N_{2Y}=6$ and 11). Such occurrences are common, because papers cited dozens of times are much more abundant than papers cited thousands of times, while there are also plenty of very small journals. Indeed, within the top 100 journals ranked by volatility (Tables 3 and 4) there are 19 journals whose top-cited paper received fewer than 50 citations and yet caused a significant volatility $\Delta f(c^*)$ that ranged from 1.6 to 4.8. High values of {\it relative volatility} $\Delta f_r(c^*)$ due to low-cited or moderately-cited papers are even more common. For 75 out of the 100 journals  in Tables 7 and 8, the top-cited paper received fewer than 10 citations and yet caused $\Delta f_r(c^*)$ to range from 90\% to 395\%.  

\begin{figure}
\centering
\includegraphics[width=0.9\linewidth]
{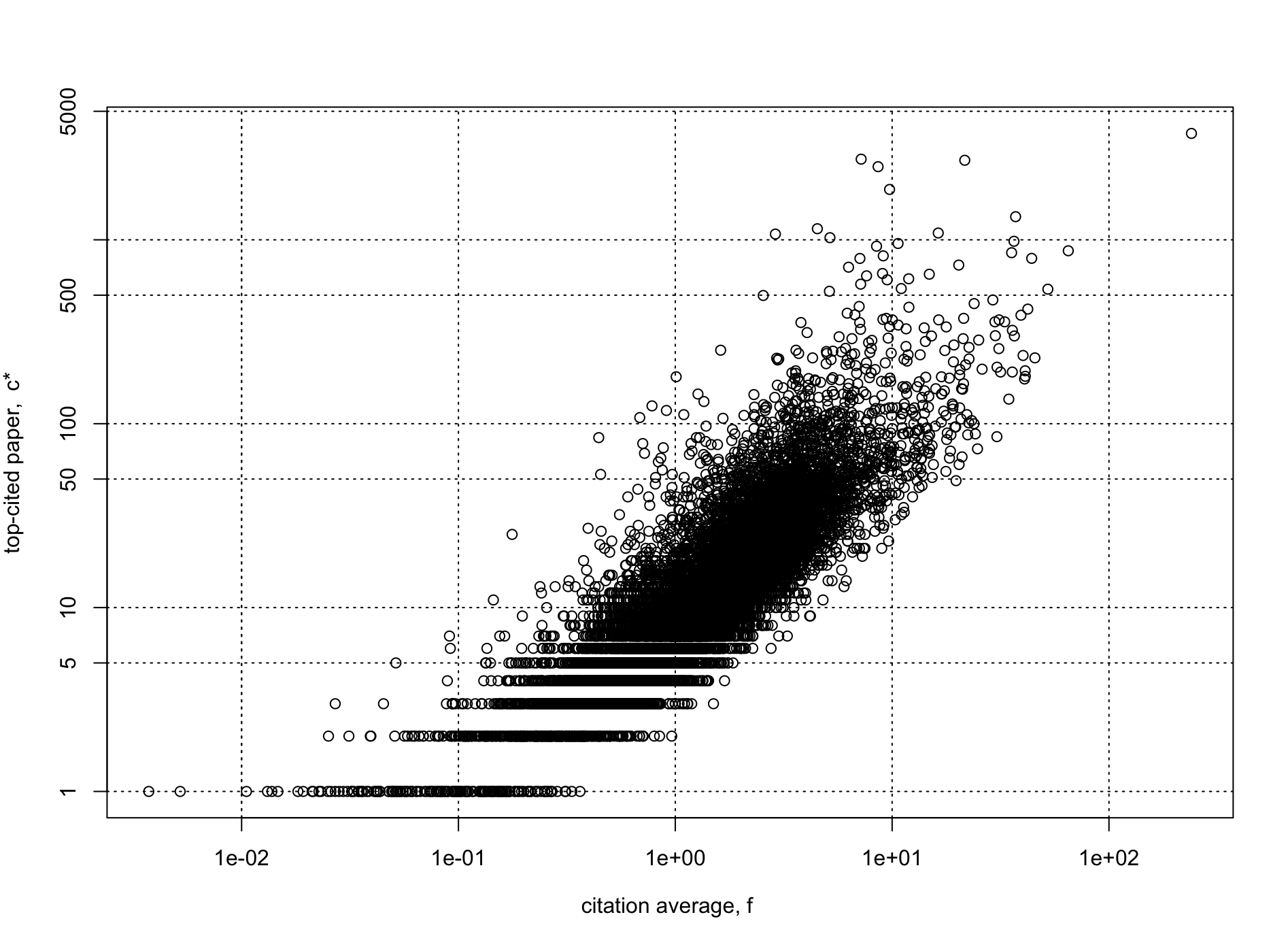}
\caption{Citations of top-cited paper, $c^*$, vs. citation average, $f$, for 11639 journals in the 2017 JCR.}
\label{fig:c*-f0}
\end{figure}

\item {\it High volatilities are observed across the IF range.} See Fig. \ref{fig:vol-f0}. For example, $\Delta f(c^*)>0.5$ for $f \sim 1-40$. High {\it relative} volatilities ($\Delta f_r(c^*) > 25\%$) are also observed across the IF spectrum. However, as expected from the Central Limit Theorem, with increasing journal size the IF approaches the global citation average $\mu=2.9$, is less sensitive to outliers and volatility drops: large bubbles ``fall'' to the bottom.

\item 
{\it The top-cited paper captures a sizable fraction of the journal's citations for journals across the IF range.} See Fig. \ref{fig:vol-f0}. The dashed line with unity slope corresponds to the situation when the top-cited paper has all the journal's citations (so that $f^*=0$ and $\Delta f(c^*) = f$). This line can never be reached in a log-log plot of data, although there are 26 journals with $f^*=0$ and another 11 journals with $f=0$, as we mentioned earlier. But note how many journals are close to that line and how they extend across the IF range. For example, 818 journals have $\Delta f_r(c^*) > 25\%$ (data points above the green line). Another example: Among the 142 journals whose top-cited paper captures more than 50\% of the journal's citations, their IF ranges from $f=0.1-21.6$ while their size ranges from $N_{2Y}=31-477$. 

\item {\it  Broadly speaking, the citation count of the top-cited paper correlates with the IF.}
See Fig.  \ref{fig:c*-f0}. But note how widely spread the highly cited papers are across journals. For example, papers with $c^* \ge 50$ appear in many journals of small-to-moderate IF, $0.5 < f < 2.5$.

\begin{table}[h]
\centering
\begin{tabular}{l l l l l l l l}
\hline
 & Journal & $\Delta f(c^*)$ & $c^*$ & $\Delta f_r(c^*)$ & $f$ & $f^*$ & $N_{2Y}$    \\
\hline
1	&	CA-CANCER J CLIN	&	68.27	&	3790	&	40\%	&	240.09	&	171.83	&	53	\\
2	&	J STAT SOFTW	&	15.80	&	2708	&	271\%	&	21.63	&	5.82	&	171	\\
3	&	LIVING REV RELATIV	&	13.67	&	87	&	273\%	&	18.67	&	5.00	&	6	\\
4	&	PSYCHOL INQ	&	8.12	&	97	&	105\%	&	15.82	&	7.70	&	11	\\
5	&	ACTA CRYSTALLOGR C	&	7.12	&	2499	&	474\%	&	8.62	&	1.50	&	351	\\
\hline
6	&	ANNU REV CONDEN MA P	&	5.67	&	209	&	35\%	&	21.82	&	16.15	&	34	\\
7	&	ACTA CRYSTALLOGR A	&	5.57	&	637	&	271\%	&	7.62	&	2.05	&	114	\\
8	&	ADV PHYS	&	4.96	&	85	&	19\%	&	30.42	&	25.45	&	12	\\
9	&	PSYCHOL SCI PUBL INT	&	4.88	&	49	&	33\%	&	19.71	&	14.83	&	7	\\
10	&	ACTA CRYSTALLOGR B	&	4.19	&	710	&	199\%	&	6.30	&	2.11	&	169	\\
\hline
11	&	NAT ENERGY	&	4.15	&	420	&	11\%	&	42.24	&	38.09	&	92	\\
12	&	INT J COMPUT VISION	&	3.83	&	656	&	74\%	&	9.03	&	5.20	&	170	\\
13	&	NAT REV MATER	&	3.65	&	228	&	9\%	&	45.55	&	41.90	&	51	\\
14	&	MOL BIOL EVOL	&	3.53	&	1879	&	57\%	&	9.73	&	6.20	&	530	\\
15	&	EPILEPSY CURR	&	3.43	&	53	&	218\%	&	5.00	&	1.57	&	15	\\
\hline
16	&	LIVING REV SOL PHYS	&	3.30	&	47	&	44\%	&	10.75	&	7.45	&	12	\\
17	&	PURE APPL CHEM	&	3.27	&	525	&	175\%	&	5.14	&	1.87	&	160	\\
18	&	PROG SOLID STATE CH	&	3.18	&	57	&	52\%	&	9.25	&	6.07	&	16	\\
19	&	PROG QUANT ELECTRON	&	3.17	&	55	&	43\%	&	10.60	&	7.43	&	15	\\
20	&	ADV OPT PHOTONICS	&	3.16	&	106	&	18\%	&	20.79	&	17.63	&	28	\\
\hline
21	&	SOLID STATE PHYS	&	3.03	&	19	&	379\%	&	3.83	&	0.80	&	6	\\
22	&	GENET MED	&	3.00	&	818	&	49\%	&	9.10	&	6.10	&	271	\\
23	&	IEEE IND ELECTRON M	&	2.93	&	89	&	42\%	&	9.86	&	6.93	&	28	\\
24	&	MATER TODAY	&	2.93	&	260	&	15\%	&	22.62	&	19.69	&	82	\\
25	&	ACTA NEUROPATHOL	&	2.93	&	650	&	25\%	&	14.84	&	11.91	&	218	\\
\hline
26	&	J METEOROL SOC JPN	&	2.81	&	247	&	130\%	&	4.98	&	2.16	&	87	\\
27	&	PROG OPTICS	&	2.67	&	32	&	103\%	&	5.27	&	2.60	&	11	\\
28	&	J BIOL ENG	&	2.58	&	103	&	101\%	&	5.13	&	2.55	&	39	\\
29	&	ANNU REV NEUROSCI	&	2.56	&	114	&	22\%	&	14.10	&	11.54	&	40	\\
30	&	ENDOCR REV	&	2.52	&	123	&	21\%	&	14.70	&	12.19	&	44	\\
\hline
31	&	CLIN MICROBIOL REV	&	2.48	&	184	&	14\%	&	20.49	&	18.02	&	67	\\
32	&	J HUM RESOUR	&	2.38	&	156	&	61\%	&	6.27	&	3.89	&	64	\\
33	&	IND ORGAN PSYCHOL-US	&	2.35	&	42	&	53\%	&	6.75	&	4.40	&	16	\\
34	&	ADV CATAL	&	2.33	&	13	&	64\%	&	6.00	&	3.67	&	4	\\
35	&	GIGASCIENCE	&	2.33	&	240	&	53\%	&	6.71	&	4.38	&	101	\\
\hline
36	&	J ACAD MARKET SCI	&	2.29	&	198	&	43\%	&	7.57	&	5.28	&	84	\\
37	&	CHINESE PHYS C	&	2.25	&	1075	&	350\%	&	2.90	&	0.64	&	477	\\
38	&	THYROID	&	2.25	&	792	&	46\%	&	7.10	&	4.85	&	350	\\
39	&	INT J CANCER	&	2.24	&	2746	&	45\%	&	7.20	&	4.96	&	1224	\\
40	&	NAT PROTOC	&	2.21	&	614	&	23\%	&	11.92	&	9.72	&	274	\\
\hline
41	&	ALDRICHIM ACTA	&	2.10	&	28	&	43\%	&	7.00	&	4.90	&	11	\\
42	&	JAMA-J AM MED ASSOC	&	1.99	&	851	&	6\%	&	35.57	&	33.57	&	410	\\
43	&	REV MOD PHYS	&	1.99	&	191	&	6\%	&	35.82	&	33.83	&	79	\\
44	&	SURF SCI REP	&	1.96	&	55	&	12\%	&	17.70	&	15.74	&	20	\\
45	&	NAT REV GENET	&	1.95	&	235	&	5\%	&	40.17	&	38.22	&	101	\\
\hline
46	&	ANNU REV ASTRON ASTR	&	1.93	&	88	&	9\%	&	24.21	&	22.27	&	34	\\
47	&	AUTOPHAGY	&	1.93	&	606	&	26\%	&	9.49	&	7.56	&	310	\\
48	&	GEOCHEM PERSPECT	&	1.92	&	9	&	144\%	&	3.25	&	1.33	&	4	\\
49	&	EUR HEART J-CARD IMG	&	1.91	&	574	&	36\%	&	7.16	&	5.25	&	298	\\
50	&	MOBILE DNA-UK	&	1.90	&	91	&	54\%	&	5.43	&	3.53	&	46	\\
\hline
\end{tabular}
\caption{Top 50 journals in volatility $\Delta f_r(c^*)$ to their top-cited paper. Publication Years = 2015--2016, Citation Year = 2017. JCR data. 11639 journals and 3,088,511 papers in data set.}
\end{table}

\begin{table}[h]
\centering
\begin{tabular}{l l l l l l l l}
\hline
 & Journal & $\Delta f(c^*)$ & $c^*$ & $\Delta f_r(c^*)$ & $f$ & $f^*$ & $N_{2Y}$    \\
\hline
51	&	NAT REV DRUG DISCOV	&	1.82	&	181	&	5\%	&	41.14	&	39.32	&	78	\\
52	&	MULTIVAR BEHAV RES	&	1.80	&	176	&	102\%	&	3.56	&	1.76	&	97	\\
53	&	EARTH SYST SCI DATA	&	1.76	&	133	&	27\%	&	8.29	&	6.54	&	72	\\
54	&	MAT SCI ENG R	&	1.65	&	60	&	9\%	&	20.36	&	18.71	&	25	\\
55	&	NAT CHEM	&	1.65	&	450	&	7\%	&	23.89	&	22.24	&	259	\\
\hline
56	&	BONE RES	&	1.64	&	89	&	16\%	&	11.92	&	10.28	&	48	\\
57	&	ANNU REV IMMUNOL	&	1.63	&	101	&	8\%	&	22.57	&	20.94	&	49	\\
58	&	NPJ COMPUT MATER	&	1.63	&	62	&	25\%	&	8.09	&	6.45	&	34	\\
59	&	PROG PART NUCL PHYS	&	1.63	&	76	&	17\%	&	10.98	&	9.35	&	41	\\
60	&	PHOTOGRAMM ENG REM S	&	1.60	&	225	&	114\%	&	3.00	&	1.40	&	140	\\
\hline
61	&	J AM SOC ECHOCARDIOG	&	1.59	&	399	&	34\%	&	6.21	&	4.62	&	248	\\
62	&	WIRES DEV BIOL	&	1.58	&	114	&	43\%	&	5.21	&	3.64	&	70	\\
63	&	J SERV RES-US	&	1.57	&	94	&	36\%	&	5.98	&	4.41	&	57	\\
64	&	NANO-MICRO LETT	&	1.57	&	137	&	29\%	&	7.02	&	5.46	&	84	\\
65	&	JAMA ONCOL	&	1.57	&	367	&	11\%	&	16.40	&	14.83	&	225	\\
\hline
66	&	WORLD PSYCHIATRY	&	1.56	&	82	&	10\%	&	17.86	&	16.29	&	42	\\
67	&	ADV APPL MECH	&	1.55	&	11	&	48\%	&	4.80	&	3.25	&	5	\\
68	&	NEURAL NETWORKS	&	1.54	&	434	&	28\%	&	7.05	&	5.51	&	279	\\
69	&	NAT REV MICROBIOL	&	1.53	&	203	&	5\%	&	30.46	&	28.94	&	114	\\
70	&	EXERC IMMUNOL REV	&	1.52	&	34	&	29\%	&	6.68	&	5.17	&	19	\\
\hline
71	&	KIDNEY INT SUPPL	&	1.51	&	23	&	82\%	&	3.36	&	1.85	&	14	\\
72	&	APPL MECH REV	&	1.51	&	55	&	29\%	&	6.73	&	5.22	&	33	\\
73	&	PROG ENERG COMBUST	&	1.51	&	73	&	6\%	&	24.76	&	23.25	&	33	\\
74	&	NEW ASTRON REV	&	1.50	&	21	&	25\%	&	7.50	&	6.00	&	10	\\
75	&	EMBO MOL MED	&	1.49	&	292	&	18\%	&	9.59	&	8.10	&	191	\\
\hline
76	&	BEHAV BRAIN SCI	&	1.47	&	46	&	31\%	&	6.21	&	4.74	&	28	\\
77	&	LANCET NEUROL	&	1.44	&	285	&	6\%	&	25.08	&	23.63	&	181	\\
78	&	NAT PHOTONICS	&	1.44	&	367	&	5\%	&	31.14	&	29.70	&	234	\\
79	&	NAT BIOTECHNOL	&	1.42	&	358	&	5\%	&	29.80	&	28.38	&	232	\\
80	&	PHYS LIFE REV	&	1.40	&	40	&	18\%	&	9.17	&	7.77	&	23	\\
\hline
81	&	NAT REV NEUROSCI	&	1.40	&	191	&	5\%	&	31.63	&	30.23	&	115	\\
82	&	J PHOTOCH PHOTOBIO C	&	1.39	&	69	&	10\%	&	14.88	&	13.49	&	40	\\
83	&	CIRCULATION	&	1.38	&	1089	&	9\%	&	16.32	&	14.93	&	776	\\
84	&	NAT REV IMMUNOL	&	1.38	&	194	&	3\%	&	41.07	&	39.69	&	112	\\
85	&	ANNU REV BIOPHYS	&	1.35	&	52	&	13\%	&	11.65	&	10.30	&	31	\\
\hline
86	&	ACTA NUMER	&	1.34	&	23	&	16\%	&	9.64	&	8.30	&	11	\\
87	&	EUR HEART J	&	1.33	&	729	&	7\%	&	20.29	&	18.95	&	532	\\
88	&	ACTA PHYS SLOVACA	&	1.33	&	13	&	133\%	&	2.33	&	1.00	&	9	\\
89	&	GENOM PROTEOM BIOINF	&	1.33	&	109	&	26\%	&	6.47	&	5.14	&	78	\\
90	&	BIOCHEM MEDICA	&	1.32	&	128	&	62\%	&	3.47	&	2.15	&	95	\\
\hline
91	&	ANNU REV FLUID MECH	&	1.32	&	70	&	10\%	&	14.67	&	13.36	&	43	\\
92	&	ANNU REV EARTH PL SC	&	1.32	&	70	&	14\%	&	10.76	&	9.44	&	46	\\
93	&	EUR J HEART FAIL	&	1.31	&	338	&	15\%	&	9.75	&	8.44	&	252	\\
94	&	ACTA ASTRONOM	&	1.31	&	65	&	57\%	&	3.60	&	2.30	&	48	\\
95	&	PSYCHOTHER PSYCHOSOM	&	1.29	&	70	&	18\%	&	8.27	&	6.98	&	49	\\
\hline
96	&	DIALOGUES HUM GEOGR	&	1.29	&	22	&	32\%	&	5.29	&	4.00	&	14	\\
97	&	LANCET INFECT DIS	&	1.28	&	336	&	8\%	&	17.76	&	16.49	&	250	\\
98	&	J ECON GROWTH	&	1.28	&	37	&	25\%	&	6.40	&	5.13	&	25	\\
99	&	J URBAN TECHNOL	&	1.27	&	61	&	91\%	&	2.66	&	1.39	&	47	\\
100	&	REV MINERAL GEOCHEM	&	1.26	&	40	&	17\%	&	8.58	&	7.32	&	26	\\
\hline
\end{tabular}
\caption{Top 51--100 journals in volatility $\Delta f(c^*)$ to their top-cited paper. Publication Years = 2015--2016, Citation Year = 2017. JCR data. 11639 journals and 3,088,511 papers in data set.}
\end{table}

\item {\it Note the parallel lines of negative slope at the bottom left corner of Fig. \ref{fig:Delta-N0}.} All these lines have slope equal to $-1$ in a log-log plot of $\Delta f(c^*)$ vs. $N_{2Y}$, a feature that is readily explained from Eq. (\ref{eq:Delta f}), whence $\Delta f(c^*) \sim (c^*-f^*)/N_{2Y}$ (since $N_{2Y} \gg 1$ usually). The offset of the parallel lines is equal to ${\rm log}(c^*-f^*)$, which for $c^* \gg f^*$ is roughly equal to ${\rm log}(c^*)$. Therefore, the $\Delta f(c^*)$ data points for all journals whose highest cited paper was cited $c^*$ times must fall on the {\it same} line, irrespective of their IF, as long as $c^* \gg f^*$. The parallel lines are therefore simply lines of increasing $c^*$ value, starting from $c^*=1, 2, 3$, etc., as we move from the bottom left to the top right of the figure. When the inequality $c^* \gg f^*$ no longer holds, a broadening of the parallel lines occurs and they overlap, exactly as we see in Fig. \ref{fig:Delta-N0}. Because of the highly skewed citation distribution of papers, the parallel lines become less populated as $c^*$ increases, i.e., for higher values of $\Delta f(c^*)$.

\begin{table}[h]
\centering
\begin{tabular}{l l l l l l l l}
\hline
Volatility, $\Delta f(c^*)$ & No.  journals & \% all journals \\
(threshold) & above threshold & \\
\hline
0.1 & 3881 & 33.3\%  \\
0.25 & 1061 & 9.1\%  \\
0.5 & 381 &	3.3\%  \\
0.75 & 221 & 1.9\% \\
\hline
1 & 140 & 1.2\%  \\
1.5 & 73 & 0.6\%  \\
2 & 41 & 0.4\%  \\
3 & 21 & 0.2\%  \\
\hline
4 & 11 & 0.1\%  \\
5 & 7 & 0.1\%  \\
10 & 3 & 0.03\%  \\
50 & 1 & 0.01\%  \\
\hline
\end{tabular}
\caption{Number of journals whose change in IF ($\Delta f(c)$) due to their highest-cited paper was {\it greater than} the threshold in the 1st column. For example, 381 journals had their IF boosted more than 0.5 points by their most cited paper. Publication Years = 2015--2016, Citation Year = 2017. Data from JCR. Total number of papers in data set is 3,090,630, published in 11639 journals.}
\end{table}

\end{enumerate}

\begin{table}[h]
\centering
\begin{tabular}{l l l l l l l l}
\hline
{\it Relative} volatility, $\Delta f_r(c^*)$ & No. journals & \% all journals \\
 (threshold) & above threshold & \\
\hline
10 \% & 	3403 & 	29.2\% \\
20 \% & 	1218 & 	10.5\% \\
25 \% & 	818 & 	7.0\%  \\
30 \% & 	592 & 	5.1\% \\
40 \% & 	387 & 	3.3\% \\
\hline
50 \% & 	231 & 	2.0\% \\
60 \% & 	174 & 	1.5\% \\
70 \% & 	140 & 	1.2\% \\
75 \% & 	127 & 	1.1\% \\
80 \% & 	124 & 	1.1\% \\
\hline
90 \% & 	101 & 	0.9\% \\
100 \% & 	50 & 	0.4\% \\
200 \% & 	14 & 	0.12\% \\
300 \% & 	5 & 	0.04\% \\
400 \% & 	1 & 	0.01\% \\
\hline
\end{tabular}
\caption{Number of journals whose {\it relative} change in IF ($\Delta f_r(c)$) due to their highest-cited paper was {\it greater than} the threshold in the 1st column. For example, 3421 journals had their IF boosted more than 10\% by their most cited paper. 
Publication Years = 2015--2016, Citation Year = 2017. Data from JCR. Total number of papers in data set is 3,090,630, published in 11639 journals.}
\end{table}

\begin{table}[h]
\centering
\begin{tabular}{l l l l l l l l}
\hline
 & Journal & $\Delta f(c^*)$ & $c^*$ & $\Delta f_r(c^*)$ & $f$ & $f^*$ & $N_{2Y}$    \\
\hline
1	&	ACTA CRYSTALLOGR C	&	7.12	&	2499	&	474	\%&	8.62	&	1.50	&	351	\\
2	&	COMPUT AIDED SURG	&	0.88	&	9	&	395	\%&	1.10	&	0.22	&	10	\\
3	&	ETIKK PRAKSIS	&	0.15	&	4	&	381	\%&	0.19	&	0.04	&	26	\\
4	&	SOLID STATE PHYS	&	3.03	&	19	&	379	\%&	3.83	&	0.80	&	6	\\
5	&	CHINESE PHYS C	&	2.25	&	1075	&	350	\%&	2.90	&	0.64	&	477	\\
\hline
6	&	LIVING REV RELATIV	&	13.67	&	87	&	273	\%&	18.67	&	5.00	&	6	\\
7	&	J STAT SOFTW	&	15.80	&	2708	&	271	\%&	21.63	&	5.82	&	171	\\
8	&	ACTA CRYSTALLOGR A	&	5.57	&	637	&	271	\%&	7.62	&	2.05	&	114	\\
9	&	AFR LINGUIST	&	0.26	&	3	&	264	\%&	0.36	&	0.10	&	11	\\
10	&	AM LAB	&	0.04	&	5	&	247	\%&	0.05	&	0.01	&	136	\\
\hline
11	&	ZKG INT	&	0.06	&	7	&	230	\%&	0.09	&	0.03	&	110	\\
12	&	EPILEPSY CURR	&	3.43	&	53	&	218	\%&	5.00	&	1.57	&	15	\\
13	&	REV INT ECON	&	1.13	&	107	&	215	\%&	1.66	&	0.53	&	94	\\
14	&	REV BRAS ORNITOL	&	0.37	&	32	&	210	\%&	0.55	&	0.18	&	85	\\
15	&	ACTA CRYSTALLOGR B	&	4.19	&	710	&	199	\%&	6.30	&	2.11	&	169	\\
\hline
16	&	REV ESP DERECHO CONS	&	0.03	&	2	&	196	\%&	0.04	&	0.01	&	76	\\
17	&	DIABETES STOFFWECH H	&	0.03	&	2	&	195	\%&	0.05	&	0.02	&	59	\\
18	&	ITINERARIO	&	0.05	&	2	&	192	\%&	0.08	&	0.03	&	37	\\
19	&	CENTAURUS	&	0.07	&	2	&	189	\%&	0.11	&	0.04	&	27	\\
20	&	HITOTSUB J ECON	&	0.07	&	2	&	189	\%&	0.11	&	0.04	&	27	\\
\hline
21	&	ACROSS LANG CULT	&	0.08	&	2	&	188	\%&	0.12	&	0.04	&	26	\\
22	&	Z ETHNOL	&	0.08	&	2	&	188	\%&	0.12	&	0.04	&	26	\\
23	&	TURK PSIKOL DERG	&	0.08	&	2	&	188	\%&	0.13	&	0.04	&	24	\\
24	&	PALAEONTOGR ABT B	&	0.31	&	6	&	184	\%&	0.47	&	0.17	&	19	\\
25	&	THEOR BIOL FORUM	&	0.11	&	2	&	183	\%&	0.17	&	0.06	&	18	\\
\hline
26	&	PURE APPL CHEM	&	3.27	&	525	&	175	\%&	5.14	&	1.87	&	160	\\
27	&	CAL COOP OCEAN FISH	&	0.60	&	16	&	167	\%&	0.96	&	0.36	&	26	\\
28	&	PROBUS	&	0.23	&	5	&	154	\%&	0.38	&	0.15	&	21	\\
29	&	NETH Q HUM RIGHTS	&	0.28	&	8	&	151	\%&	0.46	&	0.19	&	28	\\
30	&	MECH ENG	&	0.05	&	6	&	148	\%&	0.09	&	0.04	&	109	\\
\hline
31	&	GEOCHEM PERSPECT	&	1.92	&	9	&	144	\%&	3.25	&	1.33	&	4	\\
32	&	AFR NAT HIST	&	0.35	&	2	&	140	\%&	0.60	&	0.25	&	5	\\
33	&	ACTA PHYS SLOVACA	&	1.33	&	13	&	133	\%&	2.33	&	1.00	&	9	\\
34	&	J METEOROL SOC JPN	&	2.81	&	247	&	130	\%&	4.98	&	2.16	&	87	\\
35	&	ENVIRON ENG RES	&	0.78	&	80	&	129	\%&	1.38	&	0.60	&	102	\\
\hline
36	&	J RUBBER RES	&	0.11	&	4	&	127	\%&	0.20	&	0.09	&	35	\\
37	&	J AFR LANG LINGUIST	&	0.44	&	7	&	124	\%&	0.80	&	0.36	&	15	\\
38	&	FORKTAIL	&	0.15	&	6	&	115	\%&	0.28	&	0.13	&	40	\\
39	&	MEAS CONTROL-UK	&	0.38	&	19	&	114	\%&	0.71	&	0.33	&	49	\\
40	&	PHOTOGRAMM ENG REM S	&	1.60	&	225	&	114	\%&	3.00	&	1.40	&	140	\\
\hline
41	&	HEREDITAS	&	0.25	&	5	&	113	\%&	0.47	&	0.22	&	19	\\
42	&	DEV ECON	&	0.23	&	6	&	111	\%&	0.44	&	0.21	&	25	\\
43	&	EVOL EQU CONTROL THE	&	0.55	&	34	&	110	\%&	1.05	&	0.50	&	61	\\
44	&	APPL LINGUIST REV	&	0.58	&	25	&	109	\%&	1.12	&	0.54	&	42	\\
45	&	CCAMLR SCI	&	0.43	&	3	&	108	\%&	0.83	&	0.40	&	6	\\
\hline
46	&	PSYCHOL INQ	&	8.12	&	97	&	105	\%&	15.82	&	7.70	&	11	\\
47	&	JPN J MATH	&	0.38	&	6	&	105	\%&	0.73	&	0.36	&	15	\\
48	&	PROG OPTICS	&	2.67	&	32	&	103	\%&	5.27	&	2.60	&	11	\\
49	&	MULTIVAR BEHAV RES	&	1.80	&	176	&	102	\%&	3.56	&	1.76	&	97	\\
50	&	J BIOL ENG	&	2.58	&	103	&	101	\%&	5.13	&	2.55	&	39	\\
\hline
\end{tabular}
\caption{Top 50 journals in {\it relative} volatility $\Delta f_r(c^*)$ to their top-cited paper. Publication Years = 2015--2016, Citation Year = 2017. JCR data. 11639 journals and 3,088,511 papers in data set.}
\end{table}

\begin{table}[h]
\centering
\begin{tabular}{l l l l l l l l}
\hline
 & Journal & $\Delta f(c^*)$ & $c^*$ & $\Delta f_r(c^*)$ & $f$ & $f^*$ & $N_{2Y}$    \\
\hline
51	&	MANUF ENG	&	0.01	&	2	&	99	\%&	0.03	&	0.01	&	159	\\
52	&	NUCL ENG INT	&	0.01	&	1	&	99	\%&	0.01	&	0.01	&	145	\\
53	&	DEUT LEBENSM-RUNDSCH	&	0.02	&	3	&	98	\%&	0.05	&	0.02	&	133	\\
54	&	NEW REPUBLIC	&	0.01	&	1	&	98	\%&	0.02	&	0.01	&	110	\\
55	&	JCT COATINGSTECH	&	0.01	&	1	&	98	\%&	0.02	&	0.01	&	86	\\
\hline
56	&	ANDAMIOS	&	0.03	&	2	&	97	\%&	0.06	&	0.03	&	71	\\
57	&	FR CULT STUD	&	0.02	&	1	&	97	\%&	0.03	&	0.02	&	65	\\
58	&	ACTES RECH SCI SOC	&	0.05	&	3	&	97	\%&	0.09	&	0.05	&	64	\\
59	&	CEPAL REV	&	0.02	&	1	&	97	\%&	0.03	&	0.02	&	62	\\
60	&	J HELL VET MED SOC	&	0.03	&	2	&	97	\%&	0.07	&	0.04	&	58	\\
\hline
61	&	LANGAGES	&	0.05	&	3	&	96	\%&	0.11	&	0.05	&	57	\\
62	&	BER LANDWIRTSCH	&	0.02	&	1	&	96	\%&	0.04	&	0.02	&	55	\\
63	&	ANTHROPOS	&	0.02	&	1	&	96	\%&	0.04	&	0.02	&	53	\\
64	&	REV FAC AGRON LUZ	&	0.04	&	2	&	96	\%&	0.08	&	0.04	&	49	\\
65	&	AFR STUD-UK	&	0.02	&	1	&	96	\%&	0.04	&	0.02	&	49	\\
\hline
66	&	FR HIST	&	0.02	&	1	&	96	\%&	0.04	&	0.02	&	48	\\
67	&	MED GENET-BERLIN	&	0.06	&	3	&	96	\%&	0.13	&	0.07	&	47	\\
68	&	REV ROUM LINGUIST	&	0.04	&	2	&	96	\%&	0.09	&	0.04	&	47	\\
69	&	ACM T INFORM SYST SE	&	0.91	&	21	&	96	\%&	1.86	&	0.95	&	22	\\
70	&	ETHICAL PERSPECT	&	0.02	&	1	&	96	\%&	0.04	&	0.02	&	46	\\
\hline
71	&	PEDAGOG STUD	&	0.05	&	2	&	95	\%&	0.09	&	0.05	&	43	\\
72	&	TRAV GENRE SOC	&	0.02	&	1	&	95	\%&	0.05	&	0.02	&	42	\\
73	&	SECUR REGUL LAW J	&	0.02	&	1	&	95	\%&	0.05	&	0.02	&	42	\\
74	&	PULP PAP-CANADA	&	0.05	&	2	&	95	\%&	0.10	&	0.05	&	41	\\
75	&	ATLANTIS-SPAIN	&	0.02	&	1	&	95	\%&	0.05	&	0.03	&	40	\\
\hline
76	&	J HISTOTECHNOL	&	0.02	&	1	&	95	\%&	0.05	&	0.03	&	40	\\
77	&	STUD E EUR THOUGHT	&	0.02	&	1	&	95	\%&	0.05	&	0.03	&	40	\\
78	&	ETHIOP J HEALTH DEV	&	0.02	&	1	&	95	\%&	0.05	&	0.03	&	39	\\
79	&	Z ARZNEI- GEWURZPFLA	&	0.03	&	1	&	95	\%&	0.05	&	0.03	&	38	\\
80	&	INDOGER FORSCH	&	0.05	&	2	&	95	\%&	0.11	&	0.06	&	37	\\
\hline
81	&	EARTH SCI HIST	&	0.05	&	2	&	95	\%&	0.11	&	0.06	&	37	\\
82	&	CIV SZLE	&	0.03	&	1	&	95	\%&	0.05	&	0.03	&	37	\\
83	&	PSYCHOANAL STUD CHIL	&	0.05	&	2	&	94	\%&	0.11	&	0.06	&	36	\\
84	&	TIJDSCHR RECHTSGESCH	&	0.03	&	1	&	94	\%&	0.06	&	0.03	&	35	\\
85	&	TRAIT SIGNAL	&	0.03	&	1	&	94	\%&	0.06	&	0.03	&	35	\\
\hline
86	&	SOCIOL FORSKNIN	&	0.09	&	3	&	94	\%&	0.18	&	0.09	&	34	\\
87	&	MIL OPER RES	&	0.03	&	1	&	94	\%&	0.06	&	0.03	&	31	\\
88	&	J POLYNESIAN SOC	&	0.10	&	3	&	93	\%&	0.20	&	0.10	&	30	\\
89	&	AIBR-REV ANTROPOL IB	&	0.06	&	2	&	93	\%&	0.13	&	0.07	&	30	\\
90	&	AFR ASIAN STUD	&	0.06	&	2	&	93	\%&	0.13	&	0.07	&	30	\\
\hline
91	&	HIST LINGUIST	&	0.07	&	2	&	93	\%&	0.14	&	0.07	&	29	\\
92	&	RLA-REV LINGUIST TEO	&	0.18	&	5	&	93	\%&	0.37	&	0.19	&	27	\\
93	&	SCANDIA	&	0.04	&	1	&	93	\%&	0.07	&	0.04	&	27	\\
94	&	ROM J POLIT SCI	&	0.11	&	3	&	92	\%&	0.23	&	0.12	&	26	\\
95	&	GORTERIA	&	0.04	&	1	&	92	\%&	0.08	&	0.04	&	25	\\
\hline
96	&	FORSCH INGENIEURWES	&	0.32	&	8	&	92	\%&	0.67	&	0.35	&	24	\\
97	&	OBSERVATORY	&	0.04	&	1	&	91	\%&	0.09	&	0.05	&	23	\\
98	&	J URBAN TECHNOL	&	1.27	&	61	&	91	\%&	2.66	&	1.39	&	47	\\
99	&	EAST EUR COUNTRYSIDE	&	0.09	&	2	&	91	\%&	0.18	&	0.10	&	22	\\
100	&	SOCIOLOGUS	&	0.05	&	1	&	90	\%&	0.10	&	0.05	&	21	\\\hline
\end{tabular}
\caption{Top 51--100 journals in {\it relative} volatility $\Delta f_r(c^*)$ to their top-cited paper. Publication Years = 2015--2016, Citation Year = 2017. JCR data. 11639 journals and 3,088,511 papers in data set.}
\end{table}

We have studied how a journal's top-cited paper affects its IF. Could the effect work the other way around, the journal affecting citations to its papers? If such an effect were strong, the source journal would have boosted all its papers indiscriminately, and the IF-volatility would be low for all journals. Therefore, at least for the thousands of journals of high (absolute or relative) volatility, citations to the top-cited paper are mainly article-driven and not journal-driven.
 
\section{Conclusions}

Our paper has two core messages. First, we demonstrate how strongly volatile IFs are to a single or a few papers, and how frequently this occurs: Thousands of journals are seriously affected every year. Second, we demonstrate the skewed reward mechanism that affects journals' IFs disproportionately for an equally-cited paper, depending on journal size.

The above findings corroborate our earlier conclusion (Antonoyiannakis, 2018) that IFs are scale dependent in that they are particularly volatile for small journal sizes, as explained by the Central Limit Theorem. This point is pertinent for real journals because 90\% of all journals publish no more than 250 citable items annually, 
a regime where single-paper effects are at play as IFs are susceptible to sufficiently highly cited papers, of which there are thousands.

Compared to large journals, small journals have (a) {\it much} more to gain by publishing a highly-cited paper, and (b) more to lose by publishing a little-cited paper. The penalty for a zero-cited paper can be easily exceeded by the reward of a highly-cited paper. So, in terms of IF, it pays for a small journal to ``fine-tune'' its risk level: publish a few potentially groundbreaking papers, but not too many. This upper limit to how many  risky papers an elite, high-IF journal can publish before it begins to compromise its IF imparts a conservative mindset to the editor: Reject most but a few of the intellectually risky and innovative submissions, and the journal's IF can still benefit massively if some of them pay off. Such an ulterior motive---where the editor is conscious of the journal's size while assessing an individual paper at hand---makes it even harder for transformative papers to appear in elite journals.

The reliability of IF rankings (and citation averages in general) is compromised by the high IF volatility to a handful of papers, observed for thousands of journals each year. Three examples: (a) In 2017, the top cited paper of 381 journals raised their IF by 0.5; (b) 818 journals had their IF boosted by more than 25\% by their top cited paper; and (c) one in ten journals (1292 journals) had their IF boosted by more than 50\% by their top {\it three} cited papers.

Given this high sensitivity to outliers, does it make sense to compare two journals by their IF? In our view, such a comparison is at best incomplete and at worst misleading. Why incomplete? Because unless we know the underlying citation distributions and can thus ascertain that the averages (IFs) are not swayed by a few outliers, we cannot safely use IFs as {\it representatives} of both journals. And since most journals are small and have highly skewed citation distributions with outliers,  IF comparisons are more often misleading, because under these conditions the mean (average) is a poor measure of central tendency, as per standard statistical practice (De Veaux, Velleman \& Bock, 2014, pp. 57--58; Bornmann \& Mutz, 2011; Seglen, 1992; Adler, Ewing \& Taylor, 2009; Calver \& Bradley, 2009; Wall, 2009). 

So, the  volatility of IFs is not of academic but of practical interest.  It is not an exclusive feature of a few journals or a statistical anomaly that we can casually brush off, but an everyday feature inherent in citation averages, affecting thousands of journals each year. It casts serious doubt on the suitability of the IF as a {\it journal} defining quantity, and on the merits of ranking journals by IF. And it is a direct consequence of the Central Limit Theorem.

It is therefore prudent to consider novel ways of comparing journals based on more solid statistical grounds. The implications may reach much further than producing ranked lists aimed at librarians---which was the original objective of Eugene Garfield when he proposed the IF---and affect research assessment and the careers or scientists.

\section{What to do?}

Many alternatives to the IF have been proposed to date. Here we share our own recommendations for how to remedy the problem, along three lines of thought. 

\noindent 
\ding{226} {\it 1. Use metrics that are less sensitive to outliers than IFs. At a minimum, use citation medians instead of citation averages or IFs, as has been proposed by Aksnes \& Sivertsen, 2004; Rossner et al., 2007; Wall, 2009; 
Calver \& Bradley, 2009; Antonoyiannakis, 2015a, 2015b; Ioannidis \& Thombs, 2019}.

A median shows the mid-point or ``center'' of the distribution. When statisticians wish to describe the {\it typical} value of a skewed distribution, they normally report the median (De Veaux, Velleman \& Bock, 2014), together with the interquartile distance (the distance between the 1st and 3rd quartile) as a measure of the spread. Note that citation distributions are typically {\it highly} skewed, which makes the use of medians more suitable for their description. Citation medians are far less sensitive to outlier papers and much less susceptible to gaming than citation averages (IFs).
On a practical note, as of 2017, the JCR of Clarivate Analytics list the citation median per article type (research article and review article) for each journal, facilitating the wide dissemination of medians. (Cautionary note: On more occasions than we would have liked, the JCR citation medians contained errors in article type that needed correction before use.)

\noindent 
\ding{226} {\it 2. Use standardized  averages to remove the scale dependence from `bare' citation averages.} 

A `bare' average is prone to fluctuations from outliers, but the Central Limit Theorem allows us to standardize it and remove the scale dependence. So, instead of the `bare' citation average (or IF), $f$, we have proposed (Antonoyiannakis, 2018) the standardized average, or $\Phi$ index:

\begin{equation}
\Phi=\frac{f-\mu_s}{\sigma_s/\sqrt{N_{2Y}}}, \label{eq:Phi}
\end{equation}
where $\mu_s, \sigma_s$ are the global average and standard deviation of the citation distribution of all papers in the {\it subject} of the journal in question. The quantities $\mu_s$ and $\sigma_s$ need to be found for each research subject before a journal's $\Phi$ index can be calculated. For example, if we were to treat, for simplicity, all 3,088,511 papers published in all journals in 2015--2016 as belonging to a single subject, then $\mu=2.92, \sigma=8.12$, and we can use Eq. (\ref{eq:Phi}) to standardize the citation average of any journal. Here, $f$ and $N_{2Y}$ are the journal's citation average (IF) and biennial size, as usual. The $\Phi$ index is readily applicable to all citation averages, for instance in university rankings. More details will be provided in a forthcoming publication.  

\noindent 
\ding{226} {\it 3. Resist the one-size-fits-all mindset, i.e., the limitations of a single metric. Think of scholarly journals as distributions of widely varying papers, and describe them as such.} 
In line with this thinking, Larivi{\`e}re et al. (2016) suggested that journals display their full citation distribution, a recommendation adopted by several publishers so far. In a welcome development, the Clarivate Analytics JCR now display citation distributions for all journals that receive an Impact Factor. However, plots of citation distributions can be overwhelming in practice (too much information) and do not allow easy comparison across journals. So, again we turn to statistical practice and ask how  statisticians describe distributions. Typically, they use a 5-number summary of various percentiles, which is graphically displayed as a {\it box plot} and includes outlier information (De Veaux, Velleman \& Bock, 2014; Krzywinski \& Altman, 2014; Spitzer et al., 2014). We believe that use of box plots and, more generally, percentiles (Bornmann,  Leydesdorff, \& R{\" u}diger (2013), leads to responsible, informative, and practical comparisons of citation impact across journals and  other collections of papers.

\vskip 0.5cm
\noindent
{\bf Funding}

\noindent 
This research did not receive any specific grant from funding agencies in the public, commercial, or not-for-profit sectors.

\vskip 0.5cm
\noindent
{\bf Note}

\noindent
This is an extended version of an article (Antonoyiannakis, 2019) presented at the 17th International Conference of the International Society for Scientometrics and Informetrics, Rome, Italy.

\vskip 0.5cm
\noindent
{\bf Acknowledgements}

\noindent 
I am grateful to Jerry I. Dadap for stimulating discussions, and to Richard Osgood Jr. for encouragement and hospitality.
This work uses data from the Web of Science and Journal Citation Reports (2017) by Clarivate Analytics provided by Columbia University.

\vskip 0.5cm
\noindent
{\bf Author Contributions}

\noindent 
Manolis Antonoyiannakis: Conceptualization, data curation, formal analysis, methodology, writing.

\vskip 0.5cm
\noindent
{\bf Competing Interests}

\noindent
The author is an Associate Editor of {\it Physical Review B} and {\it Physical Review Research}, and a Bibliostatistics Analyst at the American Physical Society. He is also an Editorial Board member of the Metrics Toolkit, a volunteer position. He was formerly an Associate Editor of {\it Physical Review Letters} and {\it Physical Review X}. The manuscript expresses the views of the author and not of any journals, societies or institutions where he may serve.



\vskip 0.5cm
\noindent
{\bf References}
\vskip 0.5cm
%
%

\noindent 
Adler, R., Ewing, J., \& Taylor, P. (2009). A report from the International mathematical union 
\par
(IMU) in cooperation with the International council of industrial and applied mathematics 
\par 
(ICIAM) and the institute of mathematical statistics (IMS). Statistical Science, 24(1), 1--14. 
\par
https://doi.org/10.1214/09-STS285

\noindent 
Aksnes, D. W., \& Sivertsen, G. (2004). The effect of highly cited papers on national citation 
\par indicators. {\it Scientometrics, 59}, 213--224. https://doi.org/10.1023/B:SCIE.0000018529.58334.eb

\noindent 
Amin, M., \& Mabe, M. (2004). 
Impact factors: Use and abuse. 
{\it International Journal of \par
Environmental Science and Technology, 1,} 1--6.

\noindent 
Antonoyiannakis, M., \& Mitra, S. (2009). Editorial: Is PRL too large to have an `impact'? \par 
{\it Physical Review Letters}, {\it 102}, 060001. https://doi.org/10.1103/PhysRevLett.102.060001

\noindent 
Antonoyiannakis, M., Hemmelskamp, J., \& Kafatos, F. C. (2009).
The European Research \par Council  Takes Flight, {\it Cell}, {\it 136}, 805--809.
https://doi.org/10.1016/j.cell.2009.02.031 

\noindent
Antonoyiannakis, M. (2015a). Median Citation Index vs. Journal Impact Factor. APS March 
\par Meeting. Available from: http://meetings.aps.org/link/BAPS.2015.MAR.Y11.14

\noindent 
Antonoyiannakis, M. (2015b).
Editorial: Highlighting Impact and the Impact of Highlighting: \par PRB Editors' Suggestions,
{\it Physical Review B}, {\it 92}, 210001. 
\par
https://doi.org/10.1103/PhysRevB.92.210001

\noindent 
Antonoyiannakis, M. (2018). Impact Factors and the Central Limit Theorem: Why citation \par averages are scale dependent, {\it Journal of Informetrics}, {\it 12}, 1072--1088.
\par
https://doi.org/10.1016/j.joi.2018.08.011

\noindent 
Antonoyiannakis, M. (2019).
How a Single Paper Affects the Impact Factor: Implications for \par Scholarly Publishing,
{\it 
Proceedings of the 17th Conference of the International Society of \par Scientometrics and Informetrics, vol. II,} 2306--2313. 
Available from https://bit.ly/32ayyW4

\noindent 
Bornmann, L., Leydesdorff, L., \& R{\" u}diger, M. (2013). The use of percentiles and percentile rank \par classes in the analysis of bibliometric data: Opportunities and limits. {\it 
Journal of Informetrics, \par 7},  158--165.
https://doi.org/10.1016/j.joi.2012.10.001

\noindent 
Bornmann, L., \& Marx, W. (2013).
How good is research really? - Measuring the citation \par impact of publications with percentiles increases correct assessments and fair comparisons. \par 
{\it EMBO REPORTS},
{\it 14}, 226--230.
https://doi.org/10.1038/embor.2013.9

\noindent
Bornmann, L., \& Mutz, R. (2011). Further steps towards an ideal method of measuring 
\par 
citation performance: The avoidance of citation (ratio) averages in field-normalization.
\par 
{\it Journal of Informetrics, 5}(1), 228--230. 
https://doi.org/10.1016/j.joi.2010.10.009

\noindent
Calver, M. C., \& Bradley, J. S. (2009). Should we use the mean citations per paper to 
\par 
summarise a journal's impact or to rank journals in the same field? {\it Scientometrics, 
\par
81}(3), 611--615. 
https://doi.org/10.1007/s11192-008-2229-y

\noindent 
Campbell, P. (2008). 
Escape from the impact factor. 
{\it Ethics in Science and Environmental
\par
Politics, 8}, 5--7. 
https://doi.org/10.3354/esep00078 

\noindent 
Collins, F. S., Wilder, E. L., \& Zerhouni, E., (2014). NIH Roadmap/Common Fund at 10 years. \par {\it Science},  {\it 345}, (6194), 274--276.
https://doi.org/10.1126/science.1255860

\noindent 
Cope, B. \& Kalantzis, M. (2014).
Changing knowledge ecologies and the transformation of 
\par 
the scholarly journal, 
in {\it The Future of the Academic Journal (Second edition)}, 9--83.
\par
Chandos Publishing, Elsevier Limited.
https://doi.org/10.1533/9781780634647.9

\noindent 
Cornell, E., Cowley, S. Gibbs, D., Goldman, M.,  Kivelson, S.,  Quinn, H., Seestrom, S.,  Wilkins, \par J., \& Ushioda, K. (2004).
Physical Review Letters Evaluation Committee Report, \par 
http://publish.aps.org/reports/PRLReportRev.pdf 
 
\noindent 
De Veaux, R. D., Velleman, P. D., \& Bock, D. E. (2014). Stats: Data and models (3rd Edition). \par Pearson Education Limited.

\noindent 
Dimitrov, J.D., Kaveri, S.V., \& Bayry, J. (2010). Metrics: journal's impact factor skewed by a \par single paper. 
{\it Nature} {\it 466}, 179.
https://doi.org/10.1038/466179b

\noindent
Foo, J.Y.A. (2013).
Implications of a single highly cited article on a journal and its citation \par indexes: A tale of two journals.
{\it Accountability in Research, 20}, 93--106.
\par
https://doi.org/10.1080/08989621.2013.767124

\noindent 
Fortunato, S., Bergstrom, C.T., Borner, K., Evans, J.A., Helbing, D., Milojevic, S., Petersen, \par A.M., Radicchi, F., Sinatra, R., Uzzi, B., Vespignani, A., Waltman, L., Wang, D.S.,  \par \& Barabasi,  A.L. (2018).
Science of science,
{\it Science}, {\it 359}, 1007. 
\par
https://doi.org/10.1126/science.aao0185

\noindent 
Ioannidis, J. P. A., \& Thombs, B. D. (2019). A user's guide to inflated and manipulated 
\par
impact factors.  {\it European Journal of Clinical Investigation, 49}(9), 1--6. 
\par
https://doi.org/10.1111/eci.13151

\noindent 
Krzywinski, M., \& Altman, N. (2014).
Visualizing samples with box plots.
{\it Nature Methods, 11}, \par 119--120. 
https://doi.org/10.1038/nmeth.2813

\noindent
Larivi{\`e}re, V., Kiermer, V., MacCallum, C.J.,  McNutt, M., Patterson, M., Pulverer, B., \par  Swaminathan, S., Taylor, S., \& Curry, S. (2016).
A simple proposal for the publication of \par journal citation distributions,
BioRxiv, 062109. https://doi.org/10.1101/062109

\noindent 
Leydesdorff, L., Bornmann, L., \& Adams, J. (2019).
The integrated impact indicator revisited \par (I3*):
A non-parametric alternative to the journal impact factor.
{\it Scientometrics,  119}, \par 1669--1694. 
https://doi.org/10.1007/s11192-019-03099-8

\noindent 
Liu, W.S., Liu, F., Zuo, C., \& Zhu, J.W. (2018).
The effect of publishing a highly cited paper \par on a journal’s
impact factor: A case study of the Review of Particle
Physics, {\it Learned \par Publishing}  {\it 31}, 261--266.
https://doi.org/10.1002/leap.1156

\noindent 
Lyu, G., \& Shi, G. (2019).
On an approach to boosting a journal's citation potential.
\par
{\it Scientometrics, 120}, 1387--1409. 
https://doi.org/10.1007/s11192-019-03172-2

\noindent 
Milojevi{\'c}, S., Radicchi, F., \& Bar-Ilan, J. (2017). Citation success index -- An intuitive pair-wise 
\par 
journal comparison metric. Journal of Informetrics, 11(1), 223--231. 
\par
https://doi.org/10.1016/j.joi.2016.12.006

\noindent 
Moed, HF, Colledge, L, Reedijk, J, Moya-Anegon, F, Guerrero-Bote, V, Plume, A, \& Amin, M. \par (2012).
Citation-based metrics are appropriate tools in journal assessment provided that they \par are accurate and used in an informed way.
{\it Scientometrics}, {\it 92}, 367--376.
\par
https://doi.org/10.1007/s11192-012-0679-8

\noindent 
NIH News Release, (2018). 2018 NIH Director's awards for High-Risk, High-Reward \par Research program announced, \par 
https://www.nih.gov/news-events/news-releases/2018-nih-directors-awards-high-risk-high-reward-research-program-announced. 

\noindent 
Prathap, G. (2019).
Scale-dependent stratification: A skyline-shoreline scatter plot.
\par
{\it Scientometrics,  119}, 1269--1273. 
https://doi.org/10.1007/s11192-019-03038-7

\noindent
Rossner, M., Van Epps, H., \& Hill, E. (2007). 
Show me the data. 
{\it Journal of Cell Biology, 179,} \par 1091--1092.
http://jcb.rupress.org/content/179/6/1091

\noindent 
Seglen, P. O. (1992). The skewness of science. {\it Journal of the American Society for
\par 
Information Science, 43}(9), 628--638. 
\par
https://doi.org/10.1002/(SICI)1097-4571(199210)43:9<628::AID-ASI5>3.0.CO;2-0

\noindent 
Spitzer, M., Wildenhain, J., Rappsilber, J., \& Tyers, M. (2014). BoxPlotR: a web tool for \par generation of box plots.
{\it Nature Methods, 11}, 121--122.
https://doi.org/10.1038/nmeth.2811

\noindent 
Wall, H. J. (2009). Don't Get Skewed Over By Journal Rankings. {\it The B.E. Journal of Economic \par Analysis and Policy, 9}, 34. 
https://doi.org/10.2202/1935-1682.2280

\noindent 
Waltman, L., van Eck, N.J., van Leeuwen, T.N., Visser, M.S., \& van Raan, A.F.J. (2011). \par 
Towards a new crown indicator: An empirical analysis.
{\it Scientometrics}, {\it 87}, 467--481.
\par
https://doi.org/10.1007/s11192-011-0354-5

\noindent 
Wang, J., Veugelers, R., \&  Stephan, P. (2017).
Bias against novelty in science: A cautionary \par tale for users of bibliometric indicators.
{\it Research Policy},   {\it 46}, 1416--1436.
\par
https://doi.org/10.1016/j.respol.2017.06.006

\end{document}